\def\dmsol{\Delta m^2_\odot}
\def\dmatm{\Delta m^2_@}
\def\dmthtw{\Delta m^2_{32}}
\def\dmtwon{\Delta m^2_{21}}
\def\dmthon{\Delta m^2_{31}}
\def\tgatm{\tan^2\theta_@}
\def\im{{\rm Im}}
\def\tgsol{\tan^2\theta_\odot}
\def\nl{\nonumber \\}

\documentclass[superscriptaddress,aps,nofootinbib,showkeys,showpacs,preprint]{revtex4}
\usepackage{amssymb}
\usepackage{amsmath}
\usepackage{amsfonts}
\usepackage{graphicx}
\usepackage[figuresright]{rotating}
\usepackage{longtable}
\usepackage{bm}
\begin{document}

\title{Minimal Scenarios for Leptogenesis and CP Violation}

\author{G. C. Branco}
\email{gbranco@alfa.ist.utl.pt}%
\affiliation{Departamento de F{\'\i}sica and
Grupo Te{\'o}rico de F{\'\i}sica de Part{\'\i}culas, \\
Instituto Superior T{\'e}cnico, \\ Av.Rovisco Pais, 1049-001 Lisboa, Portugal}

\author{R. Gonz{\'a}lez Felipe}
\email{gonzalez@gtae3.ist.utl.pt}%
\affiliation{Departamento
de F{\'\i}sica and Grupo Te{\'o}rico de F{\'\i}sica de Part{\'\i}culas, \\
Instituto Superior T{\'e}cnico, \\ Av.Rovisco Pais, 1049-001 Lisboa, Portugal}

\author{F. R. Joaquim}
\email{filipe@gtae3.ist.utl.pt} \affiliation{Departamento de F{\'\i}sica and Grupo
Te{\'o}rico de F{\'\i}sica de Part{\'\i}culas, \\ Instituto Superior T{\'e}cnico, \\
Av.Rovisco Pais, 1049-001 Lisboa, Portugal}

\author{I. Masina}
\email{masina@spht.saclay.cea.fr}
\affiliation{Service de Physique
Th{\'e}orique, Laboratoire de la Direction des Sciences de la
Mati{\`e}re du Commissariat {\`a} l'{\'E}nergie Atomique et Unit{\'e} de
Recherche Associ{\'e}e au CNRS (URA 2306), CEA-Saclay,\\
F-91191 Gif-sur-Yvette, France}

\author{ M. N. Rebelo}
\email{rebelo@alfa.ist.utl.pt}
\affiliation{Departamento de
F{\'\i}sica and Grupo Te{\'o}rico de F{\'\i}sica de Part{\'\i}culas, \\
Instituto Superior T{\'e}cnico, \\ Av.Rovisco Pais, 1049-001
Lisboa, Portugal}

\author{C. A. Savoy}
\email{savoy@spht.saclay.cea.fr}
\affiliation{Service de Physique
Th{\'e}orique, Laboratoire de la Direction des Sciences de la
Mati{\`e}re du Commissariat
{\`a} l'{\'E}nergie Atomique et Unit{\'e} de Recherche Associ{\'e}e au
CNRS (URA 2306), CEA-Saclay,\\
F-91191 Gif-sur-Yvette, France}%

\begin{abstract}
The relation between leptogenesis and CP violation at low energies is analyzed
in detail in the framework of the minimal seesaw mechanism. Working, without
loss of generality, in a weak basis where both the charged lepton and the
right-handed Majorana mass matrices are diagonal and real, we consider a
convenient generic parametrization of the Dirac neutrino Yukawa coupling matrix
and identify the necessary condition which has to be satisfied in order to
establish a direct link between leptogenesis and CP violation at low energies.
In the context of the LMA solution of the solar neutrino problem, we present
minimal scenarios which allow for the full determination of the cosmological
baryon asymmetry and the strength of CP violation in neutrino oscillations.
Some specific realizations of these minimal scenarios are considered. The
question of the relative sign between the baryon asymmetry and CP violation at
low energies is also discussed.
\end{abstract}
\maketitle

\section{Introduction}
\label{intro}%
One of the most exciting recent developments in particle physics is the
discovery of neutrino oscillations pointed out by the Super-Kamiokande
experiment \cite{Fukuda:2001nj} and confirmed by the Sudbury Neutrino
Observatory \cite{Ahmad:2001an}. Neutrino oscillations provide evidence for
non-vanishing neutrino masses and mixings, with the novel feature that large
leptonic mixing angles are required, in contrast to what happens in the quark
sector. Indeed, the combined results from these experiments suggest that, in
addition to the large mixing angle required by the atmospheric neutrino data,
another large angle should be present in the leptonic sector. This leads to the
so-called large mixing angle (LMA) solution of the solar neutrino problem which
turns out to be presently the most favored scenario for the explanation of the
solar neutrino deficit. From a theoretical point of view, understanding the
large leptonic mixing is still an unresolved mystery for which a considerable
number of solutions have been proposed \cite{Masina:2001pp}. On the other hand,
the appearance of neutrino masses much smaller than those of charged leptons is
elegantly explained through the seesaw mechanism \cite{Yanagida:1979} which can
be implemented by extending the standard model (SM) particle content with
right-handed neutrinos. These can be easily accommodated in grand unified
theories (GUT) where they appear on equal footing with the other SM particles.

The heavy singlet neutrino states can also play an important role in cosmology,
namely, in the explanation of the observed cosmological baryon asymmetry.
During the last few years, the data collected from the acoustic peaks in the
cosmic microwave background radiation~\cite{Jungman:1995bz} has allowed to
obtain a precise measurement of the baryon asymmetry of the universe (BAU). The
MAP experiment~\cite{MAP} and the PLANCK satellite~\cite{PLANCK} planned for
the near future should further improve this result. At the present time, the
measurement of the baryon-to-entropy ratio $Y_B=n_B/s$ is
\begin{align}
\label{YBrng}%
0.7 \times 10^{-10} \lesssim Y_B\lesssim 1.0 \times 10^{-10}\,.
\end{align}

Leptogenesis is one of the most attractive mechanisms to generate the BAU. As
first suggested by Fukugita and Yanagida \cite{Fukugita:1986hr}, the key
ingredient in leptogenesis are the heavy Majorana neutrinos which, once
included in the SM, can give rise to a primordial lepton asymmetry through
their out-of-equilibrium decays. This lepton asymmetry is subsequently
reprocessed into a net baryon asymmetry by the anomalous sphaleron processes.

In spite of being attractive and successful, leptogenesis turns out to be
extremely difficult or even impossible to test experimentally in a direct way.
This difficulty is obviously related to the large masses of the heavy Majorana
neutrino singlets. Nevertheless, the joint analysis of leptogenesis and
low-energy neutrino phenomenology can be viewed as an indirect way of testing
it and here the experimental results from neutrino oscillation experiments such
as those related to the search of leptonic CP violation in the future
long-baseline neutrino experiments are extremely valuable
\cite{Lindner:2002vt}.

In this paper, we will address the question of linking the amount and sign of
the BAU to low-energy neutrino experiments, namely to the sign and strength of
the CP asymmetries measured through neutrino oscillations. Our analysis is
performed in the weak basis (WB) where the charged lepton mass matrix $m_\ell$
and the right-handed Majorana matrix $M_R$ are both real and diagonal. In this
WB, all CP-violating phases are contained in the Dirac neutrino mass matrix
$m_D$. The matrix $m_D$ is arbitrary and complex, but since three of its nine
phases can be eliminated through rephasing, one is left with six independent
physical CP-violating phases. In order to study the link between the BAU
generated through leptogenesis and CP violation at low energies, it is crucial
to use a convenient parametrization of $m_D$. We shall make use of the fact
that any arbitrary complex matrix can, without loss of generality, be written
as the product of a unitary matrix $U$ and a lower triangular matrix
$Y_\triangle$. We show that $U$ contains three phases which do not contribute
to leptogenesis, while the other three phases contained in $Y_\triangle$
contribute to both leptogenesis and low-energy CP violation. As a result, a
necessary condition for having a link between leptogenesis and low-energy CP
breaking is that the matrix $U$ contains no phases, the simplest choice being
obviously $U=\openone$. Within this class of Dirac neutrino mass matrices, we
perform a search of the minimal scenarios where not only a good fit of
low-energy neutrino data is obtained but also a link between the observed size
and sign of the BAU and the strength of CP violation observable at low energies
through neutrino oscillations can be established.
\section{\bf General Framework}
\label{generframe}%
We work in the framework of a minimal extension of the SM which consists of
adding to the standard spectrum one right-handed neutrino per generation.
Before gauge symmetry breaking, the leptonic couplings to the SM Higgs doublet
$\phi$ can be written as:
\begin{align}
{\cal L}_Y  = - Y_\nu\left(\bar{\ell}_L^{\;0},
\bar{\nu}_L^{\;0}\right)\widetilde {\phi}\,\nu_{R}^{\,0} -
Y_{\ell} \left(\bar{\ell}_L^{\;0}, \bar{\nu}_L^{\;0}\right)\phi \
\ell_{R}^{\,0} + {\rm H.c.}\,, \label{Lyuk}
\end{align}
where $\widetilde {\phi} = i \tau _2 \phi ^\ast $. After
spontaneous gauge symmetry breaking the leptonic mass terms are
given by
\begin{align}
{\cal L}_m  &= -\left[\,\bar{\nu}_L^{\;0} m_D \nu_{R}^{\,0} + \frac{1}{2}
\nu_{R}^{0\,T} C M_R \nu_{R}^{\,0}+
\bar{\ell}_L^{\;0} m_{\ell}\,\ell_R^{\,0}\,\right] + {\rm H.c.} \nonumber \\
&= - \left[\frac{1}{2}\,n_{L}^{T} C {\cal M}^* n_L+
\bar{\ell}_L^{\;0} m_{\ell}\,\ell_R^{\,0} \right] + {\rm H.c.}
\,,\label{Lmass}
\end{align}
where $m_D = v\,Y_\nu$ is the Dirac neutrino  mass matrix with $v=\langle
\phi^{\,0} \rangle /\sqrt{2} \simeq 174\,$GeV, $M_R$ and $m_\ell=v\,Y_{\ell}$
denote the right-handed Majorana neutrino and charged lepton mass matrices,
respectively, and $n_L= ({\nu}_{L}^0, {(\nu_R^0)}^c)$. Among all the terms,
only the right-handed neutrino Majorana mass term is SU(2) $\times$ U(1)
invariant and, as a result, the typical scale of $M_R$ can be much above  the
electroweak symmetry breaking scale $v$, thus leading to naturally small
left-handed Majorana neutrino masses of the order $m^2_D/M_R$ through the
seesaw mechanism. In terms of weak-basis eigenstates the leptonic charged
current interactions are given by:
\begin{align}
{\cal L}_W  = -\frac{g}{\sqrt 2}W^{+}_{\mu}\,\bar{\nu}_L^{\;0} \,
\gamma ^{\mu}\,\ell_L^{\,0} + {\rm H.c.}\,. \label{Lcc1}
\end{align}
It is clear from Eqs. (\ref{Lmass}) and (\ref{Lcc1}) that it is possible to
choose, without loss of generality, a weak basis (WB) where both $m_\ell$ and
$M_R$ are diagonal, real and positive. Note that in this WB, $m_D$ is a general
complex matrix which contains all the information on CP-violating phases. Since
in the present framework there is no $\Delta L=2$ mass term of the form
$\frac{1}{2} \nu_{L}^{0\,T} C M_L \nu_{L}^0$, the total number of CP-violating
phases for $n$ generations is given by $n(n -1)$ \cite{Endoh:2000hc} which are
all contained in $m_D$ in this special weak basis\footnote{The counting of
independent CP-violating phases for the general case, where besides $m_D$ and
$M_R$ there is also a left-handed Majorana mass term at tree level has been
discussed in Ref.~\cite{Branco:gr}.}.

We recall that the full $6 \times 6$ neutrino mass matrix $\mathcal{M}$ is
diagonalized via the transformation:
\begin{align}
V^T {\cal M}^* V = \cal D , \label{Mnudi}
\end{align}
where ${\cal D} ={\rm diag} (m_1, m_2, m_3, M_1, M_2, M_3)$, with $m_i$ and
$M_i$ denoting the physical masses of the light and heavy Majorana neutrinos,
respectively. It is convenient to write $V$ and $\cal D$ in the following form,
together with the definition of $\cal M$ :
\begin{align}
V= \left (\begin{array}{cc}
K & Q \\
S & T \end{array}\right) \;\;,\;\;{\cal D}=\left(\begin{array}{cc}
d_\nu & 0 \\
0 & D_R \end{array}\right)\;\;,\;\;{\cal M}= \left (\begin{array}{cc}
0 & m_D \\
m_D^{\,T} & M_R \end{array}\right).
\end{align}
From Eq. (\ref{Mnudi}) one obtains, to an excellent approximation, the seesaw
formula:
\begin{align}
d_\nu\simeq -K^\dagger\, m_D\, M_R^{-1}\, m_D^{\,T}\, K^* \equiv
K^\dagger\,\mathcal{M}_\nu\, K^*\,, \label{ssaw}
\end{align}
where $\mathcal{M}_\nu$ is the usual light neutrino effective mass
matrix. The leptonic charged-current interactions are given by
\begin{align}
- \frac{g}{\sqrt{2}} \left( \bar{\ell}_{L} \,\gamma_{\mu} \, K
{\nu}_{L} + \bar{\ell}_{L}\, \gamma_{\mu}\, Q\, N_{L} \right)
W^{\mu} +{\rm H.c.}\,, \label{Lcc2}
\end{align}
where $\nu_i$ and  $N_i$ denote the light and heavy neutrino mass eigenstates,
respectively. The matrix $K$ which contains all information on mixing and CP
violation at low energies can then be parametrized, after eliminating the
unphysical phases, by $K= { U_{\delta}} P$ with $P ={\rm diag}(1,
e^{i\,\alpha},e^{i\,\beta})$ ($\alpha$ and $\beta$ are Majorana phases) and
$U_{\delta}$ a unitary matrix which contains only one (Dirac-type) phase
$\delta$. In the limit where the heavy neutrinos exactly decouple from the
theory, the matrix $K$ is usually referred as the
Pontecorvo-Maki-Nakagawa-Sakata mixing matrix, which from now on we shall
denote as $U_\nu$.

\medskip \bigskip
\textbf{CP violation in neutrino oscillations}
\medskip \bigskip

It has been shown \cite{Branco:2001pq} that the strength of CP violation at low
energies, observable for example through neutrino oscillations, can be obtained
from the following low-energy WB invariant:
\begin{align}
{\cal T}_{CP} = {\rm Tr}\left[\,\mathcal{H}_{\nu}, H_\ell
\,\right]^3=6\,i \,\Delta_{21}\,\Delta_{32}\,\Delta_{31}\,{\rm Im}
\left[\,
(\mathcal{H}_{\nu})_{12}(\mathcal{H}_{\nu})_{23}(\mathcal{H}_{\nu})_{31}\,
\right]\,, \label{TCP}
\end{align}
where $\mathcal{H}_{\nu}=\mathcal{M}_\nu\,\mathcal{M}_\nu^{\dag}$,
$H_\ell=m_\ell\,{m_\ell}^{\dagger}$ and $\Delta_{21}=({m_{\mu}}^2-{m_e}^2)$
with analogous expressions for $\Delta_{31}$, $\Delta_{32}$. This relation can
be computed in any weak basis. The low-energy invariant (\ref{TCP}) is
sensitive to the Dirac-type phase $\delta$ and vanishes for $\delta=0$. On the
other hand, it does not depend on the Majorana phases $\alpha$ and $\beta$
appearing in the leptonic mixing matrix. The quantity ${\cal T}_{CP}$ can be
fully written in terms of physical observables once
\begin{align}
{\rm Im} \left[\, (\mathcal{H}_{\nu})_{12}(\mathcal{H}_{\nu})_{23}
(\mathcal{H}_{\nu})_{31}\, \right] = - \dmtwon\,\dmthon\,\dmthtw\,{\cal
J}_{CP}\,, \label{ImHHH}
\end{align}
where the $\Delta m_{ij}^2$'s are the usual light neutrino mass squared
differences and ${\cal J}_{CP}$ is the imaginary part of an invariant quartet
appearing in the difference of the CP-conjugated neutrino oscillation
probabilities $P(\nu_e\rightarrow\nu_\mu)-P(\bar{\nu}_e\rightarrow
\bar{\nu}_\mu)$. One can easily get:
\begin{align}
{\cal J}_{CP} &\equiv {\rm Im}\left[\,(U_\nu)_{11} (U_\nu)_{22}
(U_\nu)_{12}^\ast (U_\nu)_{21}^\ast\,\right] \nonumber \\
&= \frac{1}{8} \sin(2\,\theta_{12}) \sin(2\,\theta_{13}) \sin(2\,\theta_{23})
\cos(\theta_{13})\sin \delta\,, \label{Jgen1}
\end{align}
where the $\theta_{ij}$ are the mixing angles appearing in the standard
parametrization adopted in \cite{Hagiwara:pw}. Alternatively, one can use
Eq.~(\ref{ImHHH}) and write:
\begin{align} {\cal J}_{CP}=-\frac{{\rm Im}\left[\,
(\mathcal{H}_{\nu})_{12}(\mathcal{H}_{\nu})_{23}(\mathcal{H}_{\nu})_{31}\,
\right]}{\dmtwon\,\dmthon\,\dmthtw}\,. \label{Jfin}
\end{align}
This expression has the advantage of allowing the computation of the low-energy
CP invariant without resorting to the mixing matrix $U_\nu$.

It is also possible to write WB invariants useful to leptogenesis
\cite{Branco:2001pq} as well as WB invariant conditions for CP conservation in
the leptonic sector relevant in specific frameworks
\cite{Branco:gr,Branco:1999bw}.

\section{{\bf CP asymmetries in heavy Majorana neutrino decays}}
\label{CPasymmetries}%
The starting point in leptogenesis scenarios is the $CP$ asymmetry generated
through the interference between tree-level and one-loop heavy Majorana
neutrino decay diagrams. In the simplest extension of the SM, such diagrams
correspond to the decay of the Majorana neutrino into a lepton and a Higgs
boson. Considering the decay of one heavy Majorana neutrino $N_j$, this
asymmetry is given by:
\begin{align} \label{epsin}
\varepsilon_j=\frac{\Gamma\,(N_j \rightarrow \ell\,\phi)-\Gamma
\,(N_j \rightarrow \bar{\ell}\,\phi^{\,\dag})}{\Gamma\,(N_j
\rightarrow \ell\,\phi)+\Gamma\,(N_j \rightarrow
\bar{\ell}\,\phi^{\,\dag})}\ .
\end{align}
In terms of the Dirac neutrino Yukawa couplings the CP asymmetry (\ref{epsin})
is \cite{Covi:1996wh}:
\begin{align}
\label{epsj1} \varepsilon_j=\frac{1}{8\pi(Y_\nu^{\dag}
Y_\nu^{})_{jj}} \sum_{k\neq j}\,\im[\,(Y_\nu^{\dag}\,Y_\nu^{})_
{jk}^2\,]\,f\!\left (\frac{M_k^{\,2}}{M_j^{\,2}}\right)\,,
\end{align}
where the index $j$ is not summed over in $(Y_\nu^{\dag} Y_\nu^{})_{jj}\,$. The
loop function $f(x)$ includes the one-loop vertex and self-energy corrections
to the heavy neutrino decay amplitudes,
\begin{align}
\label{f} f(x)=\sqrt{x} \left[\,(1+x)\ln \left(\frac{x}
{1+x}\right)+\frac{2-x}{1-x}\,\right]\,.
\end{align}
From Eq.~(\ref{epsj1}) it can be readily seen that the CP asymmetries are only
sensitive to the CP-violating phases appearing in $Y_\nu^\dagger Y_\nu^{}$ (or
equivalently in $m_D^\dagger m_D^{}$) in the WB where $M_R$ and $m_\ell$ are
diagonal.

\subsection{Hierarchical case: $\bm{M_1 < M_2 \ll M_3}$ }

In the hierarchical case $M_1 < M_2 \ll M_3$, only the decay of the lightest
heavy neutrino $N_1$ is relevant for leptogenesis, provided the interactions of
$N_1$ are in thermal equilibrium at the time $N_{2,3}$ decay, so that the
asymmetries produced by the latter are erased before $N_1$ decays. In this
situation, it is sufficient to take into account the CP asymmetry
$\varepsilon_1$. Since in the limit $x \gg 1$ the function $f(x)$ can be
approximated by\footnote{This approximation can be reasonably used for $x
\gtrsim 15$.} $f(x)\simeq -3/(2\sqrt{x})$, we have from Eq.~(\ref{epsj1})
\begin{align}
\label{eps1} \varepsilon_1=-\frac{3}{16\pi(Y_\nu^{\dag}
Y_\nu^{})_{11}} \sum_{k=2,3}\,\im[\,(Y_\nu^{\dag}\,Y_\nu^{})_
{1k}^2\,]\,\frac{M_1}{M_k}\,,
\end{align}
which can be recast in the form \cite{Buchmuller:2000nd}
\begin{align}
\varepsilon_1 \simeq -\frac{3\,M_1}{16\,\pi}\frac{ {\rm
Im}\left[\,Y_\nu^\dagger\,Y_\nu^{}\,D_R^{-1}\,Y_\nu^T\,Y_\nu^\ast\,\right]
_{11}}{(Y_\nu^{\dag}Y_\nu^{})_{11}}=\frac{3\,M_1}{16\,\pi\,v^2}\frac{
{\rm Im}\left[\,Y_\nu^\dagger\,\mathcal{M}_{\nu}\,
Y_\nu^\ast\,\right]_{11}} {(Y_\nu^{\dag} Y_\nu^{})_{11}}\,,
\label{ep1Mn}
\end{align}
using the seesaw relation given in Eq.~(\ref{ssaw}).
\subsection{Two-fold quasi-degeneracy: $\bm{M_1 \simeq M_2 \ll M_3}$}
\label{degenRH}%

In the context of thermal leptogenesis, when the observed baryon asymmetry is
generated through the decays of the lightest heavy Majorana neutrino $N_1$,
there exists an upper bound on the CP asymmetry $\varepsilon_1$ which directly
depends on the mass of the lightest neutrino $M_1$ \cite{Davidson:2002qv}. In
turn, such a bound implies a lower bound on the lightest mass $M_1$, typically
$M_1 \gtrsim 10^{8}$~GeV. The latter bound\footnote{A more stringent
constraint, $M_1 \gtrsim 10^{10}$~GeV, is obtained in \cite{Ellis:2002eh}.} is
however barely compatible with the reheating temperature bound $T_R \lesssim
10^{8} \sim 10^9$~GeV required in several supergravity models in order to avoid
a gravitino overproduction \cite{Ellis:1984eq}. To overcome this
problem\footnote{Another possible solution to the gravitino problem is to
consider non-thermal production mechanisms \cite{Giudice:1999fb}. Since in
these cases the condition $M_1 < T_R$ is not required, the gravitino problem is
easily avoided once heavy particles can be created with a relatively low
reheating temperature without threatening big bang nucleosynthesis.} one can
consider, for instance, the decays of two heavy neutrinos which are
quasi-degenerate in mass, $M_1 \simeq M_2$. In this case, the CP asymmetries
$\varepsilon_i$ are enhanced due to self-energy contributions
\cite{Flanz:1994yx} and the required baryon asymmetry can be produced by
right-handed heavy neutrinos with moderate masses $M_1 \simeq M_2 \lesssim
10^8$~GeV. Moreover, it has been shown that in the presence of small Dirac-type
leptonic mixing at high energies and GUT-inspired Dirac neutrino Yukawa
couplings, the heavy Majorana neutrino degeneracy is compatible with the LMA
solar solution \cite{GonzalezFelipe:2001kr}.

Let us assume that the heavy Majorana neutrinos $N_1$ and $N_2$ are
quasi-degenerate. It is useful to define the parameter $\delta_N$ which
represents the degree of degeneracy between the masses $M_1$ and $M_2$ as
\begin{align}
\label{deltN} \delta_N=\frac{M_2}{M_1}-1\,.
\end{align}
Since $M_1 \simeq M_2\ $, we expect $\delta_N \ll 1$. For the perturbative
approach to remain valid, the tree-level decay width $\Gamma_i$ for each of the
heavy Majorana neutrinos must be much smaller than the mass difference between
them. This is translated into the relations
\begin{align}
\label{crit1} \Gamma_i = \frac{(H_\nu)_{ii}\,M_i}{8\,\pi} \ll
M_2-M_1 =\delta_N\,M_1\,, \quad i=1,2\,,
\end{align}
where $H_\nu=Y_\nu^\dagger\,Y_\nu$. From this equation we can find the
following lower bound for $\delta_N$:
\begin{align}
\label{dNlim} \delta_N \gg {\rm
max}\left\{\frac{(H_\nu)_{ii}\,M_i}{8\,\pi\,M_1}\right\}_{i=1,2}\,.
\end{align}
Assuming that this criterium is verified, the CP asymmetries $\varepsilon_i$
can be obtained combining Eqs.~(\ref{epsj1}) and (\ref{deltN}). We find
\begin{align}
\label{e1e2d}
\varepsilon_1&=-\frac{1}{8\,\pi\,(H_\nu)_{11}}\left\{{\rm
Im}\,[\,(H_\nu)_{21}^2\,]\,f[\,(1+\delta_N)^2\,]-\frac{3}{2}\,{\rm
Im} [\,(H_\nu)_{31}^2\,]\, \frac{M_1}{M_3} \right\}\,,\nl
\varepsilon_2&=-\frac{1}{8\,\pi\,(H_\nu)_{22}}\left\{{\rm
Im}\,[\,(H_\nu)_{12}^2\,]\,f[\,(1+\delta_N)^{-2}\,]-\frac{3}{2}\,{\rm
Im} [\,(H_\nu)_{32}^2\,]\, \frac{M_2}{M_3}\right\}\,.
\end{align}
Taking into account that for $\delta_N \ll 1$ the function
$f[\,(1+\delta_N)^{\pm 2}\,]$ can be approximated by\footnote{For $\delta_N <
10^{-2}$ the error associated to this approximation is less than 1\%.}:
\begin{align}
\label{fapp2} f[\,(1+\delta_N)^{\pm 2}\,]\simeq \mp
\frac{1}{2\,\delta_N}\,,
\end{align}
we obtain:
\begin{align}\label{e12d2}
\varepsilon_j&=\frac{1}{16\,\pi\,(H_\nu)_{jj}}\left[\frac{{\rm
Im}\,[\,(H_\nu)_{21}^2\,]}{\delta_N}+ 3\,{\rm Im}
[\,(H_\nu)_{3j}^2\,]\, \frac{M_j}{M_3}\right]\;\;,\;\; j=1,2\,.
\end{align}
Typically, the term proportional to $M_j/M_3$ can be neglected and in this case
$\varepsilon_1$ and $\varepsilon_2$ have the same sign. This aspect turns out
to be relevant for the discussion on the relative sign between the BAU and
low-energy leptonic CP violation.
\section{On the connection between leptogenesis and low-energy CP violation}
\label{parametrizations}%
In this section we analyze the possible link between CP violation at low
energies, measurable for example through neutrino oscillations, and
leptogenesis. The possibility of such a connection has been previously analyzed
in the literature \cite{Branco:2001pq,Branco:2002kt,Frampton:2002qc}.
Nevertheless, we find it worthwhile presenting here a thorough discussion on
the subject. In particular, we will address the following questions:
\begin{itemize}
\item{If the strength of CP violation at low energies in neutrino oscillations
is measured, what can one infer about the viability
or non-viability of leptogenesis?}%
\item{From the sign of the BAU, can one predict the sign of the CP asymmetries
at low energies, namely the sign of $\mathcal{J}_{CP}$?}
\end{itemize}
We will show that having an explicit parametrization of $m_D$ (or equivalently
of $Y_\nu=m_D/v$) is crucial not only to determine which phases are responsible
for leptogenesis and which ones are relevant for leptonic CP violation at low
energies, but also to analyze the relationship between these two phenomena.

From the available neutrino oscillation data, one obtains some information on
the effective neutrino mass matrix $\mathcal{M}_\nu$ which can be decomposed in
the following way:
\begin{align}
U_{\nu}\,d_\nu\,U_{\nu}^{\,T}=\mathcal{M}_{\nu} \equiv
L\,L^T\;,\;L\equiv i\,m_D\,D_R^{-1/2}\,. \label{MnuLL}
\end{align}
The extraction of $L$ from $\mathcal{M}_\nu$ suffers from an intrinsic
ambiguity \cite{Casas:2001sr} in the sense that, given a particular solution
$L_0$ of Eq.~(\ref{MnuLL}), the matrix $L=L_0\,R$ will also satisfy this
equation, provided that $R$ is an arbitrary orthogonal complex matrix, $R \in
O(3, \mathbb{C})$, i.e. $R\,R^T=\openone$. It is useful to take as a reference
solution $L_0 \equiv U_\nu\,d_\nu^{1/2}$, so that:
\begin{align}
\label{L2} L \equiv U_\nu\,d_\nu^{1/2} R\,.
\end{align}
Since three of the phases of $m_D$ can be eliminated, the matrix $L$ has 15
independent parameters. The parametrization of $L$ given in Eq.~(\ref{L2}) has
the interesting feature that all its parameters are conveniently distributed
among $U_\nu$, $d_\nu$ and $R$, which contain 6 (3 angles + 3 phases), 3 and 6
(3 angles + 3 phases) independent parameters, respectively. Of the 18
parameters present in the Lagrangian of the fundamental theory described by
$m_D$ and $D_R$, only 9 appear at low energy in $\mathcal{M}_\nu$ through the
seesaw mechanism. To further disentangle $m_D$ from $D_R$ in $L$, one needs the
3 remaining inputs, namely the three heavy Majorana masses of $D_R$. As for the
meaning of the information encoded in $R$, it turns out that the pattern of
this matrix has a suggestive interpretation in terms of the different r\^oles
played by the heavy neutrinos in the seesaw mechanism. In fact, $R$ can be
viewed as a dominance matrix \cite{Lavignac:2002gf} since it gives the weights
of each heavy Majorana neutrino in the determination of the different light
neutrino masses $m_i$ \cite{Masina:2002qh}. The fact that $R_{ij}^2$ are
weights for $m_i$ is quite obvious due to the orthogonality of $R$:
\begin{align}
m_i  = \sum_j  m_i R_{ij}^{\,2}\,.
\end{align}
On the other hand, since $U_\nu^{\,\dagger}\,m_D =-i\, d_\nu^{1/2}R\,
D_R^{1/2}$, the single contribution $m_i R_{ij}^{\,2}$ is also given by:
\begin{align}
m_i R_{ij}^{\,2}= -\frac{(U_\nu^\dagger m_D)^2_{ij}}{M_j} \equiv
\frac{X_{ij}}{M_j}\,.
\end{align}
Therefore, once $U_\nu$ is fixed, each weight $R_{ij}^2$ just depends on the
mass $M_j$ of the j-th heavy Majorana neutrino and on its couplings with the
left-handed neutrinos $(m_D)_{kj}$. Thus, the contribution of each heavy
neutrino to $m_i$ is well defined and expressed by the weight ${\rm
Re}(R_{ij}^2)$. One may roughly say that the heavy Majorana neutrino with mass
$M_j$ dominates\footnote{See for instance Refs.~\cite{Smirnov:af,
Altarelli:1999dg} for other approaches to the dominance mechanism.} in $m_i$ if
\begin{align}
\frac{|{\rm Re}\,(X_{ij})|}{M_j} \gg \frac{|{\rm
Re}\,(X_{ik})|}{M_k}\;\;,\;\;k\ne j
\end{align}
which implies $|{\rm Re}(R_{ij}^{\,2})| \gg |{\rm Re}(R_{ik}^{\,2})|\,$. So
that, if one of the heavy Majorana neutrino neutrino gives the dominant
contribution to one of the masses $m_i$, this information is encoded in the
structure of $R$. The interpretation in terms of weights is straightforward for
the rotational part of $R$. However, one has to be careful because in the
presence of the three boosts (controlled by the three phases) the weights
$R_{ij}^2$ are not necessarily real and positive. Although this situation is
more subtle, the above dominance arguments still hold.

Coming back to the connection between leptogenesis and low-energy data, it is
important to note that $U_\nu$ does not appear in the relevant combination for
leptogenesis $Y_\nu^{\dag}Y_\nu^{}$, in the same way as $R$ does not appear in
$\mathcal{M}_\nu$. Indeed, one has:
\begin{align}
 m_D^\dagger m_D^{} = D_R^{1/2}\,R^{\dagger}\,d_\nu\,R\,D_R^{1/2}
 \,.\label{HHdag}
\end{align}
From the above discussion, it follows that it is possible to write $m_D$ in the
form $m_D=-i\,U_\nu\,d_{\nu}^{1/2}R\,D_R^{1/2}$ in such a way that leptogenesis
and the low-energy neutrino data (contained in $\mathcal{M}_\nu$) depend on two
independent sets of CP-violating phases, respectively those in $R$ and those in
$U_{\nu}$. In particular, one may have viable leptogenesis even in the limit
where there are no CP-violating phases (neither Dirac nor Majorana) in $U_\nu$
and hence, no CP violation at low energies \cite{Rebelo:2002wj}. Therefore, in
general it is not possible to establish a link between low-energy CP violation
and leptogenesis. This connection is model dependent: it can be drawn only by
specifying a particular \emph{ansatz} for the fundamental parameters of the
seesaw, $m_D$ and $D_R$, as will be done in the following sections.

The relevance of the matrix $R$ for leptogenesis can be rendered even more
explicit \cite{Masina:2002qh} by rewriting the $\varepsilon_1$ asymmetry by
means of Eq.~(\ref{HHdag}) and defining $R_{ij}= |R_{ij}| e^{i
\varphi_{ij}/2}$, $\dmsol \equiv\dmtwon$ and $\dmatm \equiv \dmthtw$. In the
case of hierarchical heavy Majorana neutrinos, say $M_1 \ll M_2 \ll M_3$ one
obtains
\begin{align}
\varepsilon_1 \simeq \frac{3}{16\pi} \frac{M_1}{v^2} \frac {
\dmatm |R_{31}|^2 \sin \varphi_{31}  - \dmsol  |R_{11}|^2
\sin\varphi_{11}} { m_1|R_{11}|^2 +m_2|R_{21}|^2
+m_3|R_{31}|^2}\,, \label{edb}
\end{align}
and we recover what one would have expected by intuition, namely that the
physical quantities involved in determining $\varepsilon_1$ are just $M_1$, the
spectrum of the light neutrinos, $m_i$, and the first column of $R$, which
expresses the composition of the lightest heavy Majorana neutrino in terms of
the light neutrino masses $m_i$. In the case $M_1 \simeq M_2 \ll M_3$, similar
expressions hold,
\begin{align}
\varepsilon_j \simeq \frac{1}{16 \pi v^2} \frac{M_1 M_2}{M_2-M_1}
\frac{{\rm Im}[(R^\dagger d_\nu R)_{21}^2\,]} {(R^\dagger d_\nu
R)_{jj}}\,, \quad j=1,2 \,.
\end{align}
where now also the mass $M_2$ and the second column of $R$ are involved. A
detailed study of the relevance of the matrix $R$ for leptogenesis is under way
\cite{inprep}.

As stressed before, different \emph{ans\"atze} for $R$ have no direct impact on
CP violation at low energy; the impact is in a sense indirect because $R$
specifies if dominance of some heavy Majorana neutrino is at work in the seesaw
mechanism \cite{Lavignac:2002gf}.

In conclusion, the link between leptogenesis and low-energy CP violation can
only be established in the framework of specific \emph{ans\"atze} for the
leptonic mass terms of the Lagrangian. We shall derive a necessary condition
for such a link to exist. In order to obtain this connection, it is convenient
to use a triangular parametrization for $m_D$, which we describe next.

\medskip \bigskip
\textbf{Triangular parametrization}
\medskip \bigskip

It can be easily shown that any arbitrary complex matrix can be written as the
product of a unitary matrix U with a lower triangular matrix $Y_{\triangle}$.
In particular, the Dirac neutrino mass matrix can be written as:
\begin{align}
m_D = v\,U\,Y_{\triangle}\,, \label{mDtri}
\end{align}
with $Y_{\triangle}$ of the form:
\begin{align}
Y_{\triangle}= \left(\begin{array}{ccc}
y_{11} & 0 & 0 \\
y_{21}\,e^{i\,\phi_{21}} & y_{22} & 0 \\
y_{31}\,e^{i\,\phi_{31}} & y_{32}\,e^{i\,\phi_{32}} & y_{33}
\end{array}
\right)\,, \label{Ytri1}
\end{align}
where $y_{ij}$ are real positive numbers. Since $U$ is unitary, in general it
contains six phases. However, three of these phases can be rephased away by a
simultaneous phase transformation on ${\nu}_{L}^{\,0}$, $\ell_{L}^{\,0}$, which
leaves the leptonic charged current invariant. Under this transformation, $m_D
\rightarrow P_{\xi} m_D $, with $P_{\xi}={\rm diag} \left(e^{i \, \xi_1},e^{i
\, \xi_2},e^{i \, \xi_3} \right)$. Furthermore, $Y_{\triangle}$ defined in
Eq.~(\ref{Ytri1}) can be written as:
\begin{align}
Y_{\triangle}= {P_{\beta}^\dagger}\ {\hat Y_{\triangle}}\ P_{\beta}\,,
\label{Ytri2}
\end{align}
where $P_\beta ={\rm diag} (1, e^{i \, \beta_1}, e^{i \, \beta_2})$ with
$\beta_1=-\phi_{21}$, $\beta_2=-\phi_{31}$ and
\begin{align}
{\hat Y_{\triangle}}= \left(\begin{array}{ccc}
y_{11} & 0 & 0 \\
y_{21}  & y_{22} & 0 \\
y_{31}  & y_{32}\,e^{i \, \sigma} & y_{33}
\end{array}
\right) \,,\label{Ytri3}
\end{align}
with $\sigma=\phi_{32}- \phi_{31}+ \phi_{21}$. It follows then from
Eqs.~(\ref{mDtri})~and~(\ref{Ytri2}) that the matrix $m_D$ can be decomposed in
the form
\begin{align}
m_D=v\,U_{\rho}\,P_{\alpha}\,{\hat Y_{\triangle}}\,P_{\beta}\,,
\label{mDdec}
\end{align}
where $P_\alpha ={\rm diag} (1, e^{i \, \alpha_1}, e^{i \, \alpha_2})$ and
$U_\rho$ is a unitary matrix containing only one phase $\rho$. Therefore, in
the WB where $m_\ell$ and $M_R$ are diagonal and real, the phases $\rho$,
$\alpha_1$, $\alpha_2$, $\sigma$, $\beta_{1}$ and $\beta_{2}$ are the only
physical phases characterizing CP violation in the leptonic sector. The phases
relevant for leptogenesis are those contained in $m_D^{\,\dagger} m_D^{}$. From
Eqs.~(\ref{Ytri2})-(\ref{mDdec}) we conclude that these phases are $\sigma$,
$\beta_{1}$ and $\beta_{2}$, which are linear combinations of the phases
$\phi_{ij}$. On the other hand, all the six phases of $m_D$ contribute to the
three phases of the effective neutrino mass matrix at low energies
\cite{Branco:2001pq} which in turn controls CP violation in neutrino
oscillations. Since the phases $\alpha_1$, $\alpha_2$ and $\rho$ do not
contribute to leptogenesis, it is clear that a necessary condition for a direct
link between leptogenesis and low-energy CP violation to exist is the
requirement that the matrix $U$ in Eq. (\ref{mDtri}) contains no CP-violating
phases. Note that, although the above condition was derived in a specific WB
and using the parametrization of Eq. (\ref{mDtri}), it can be applied to any
model. This is due to the fact that starting from arbitrary leptonic mass
matrices, one can always make WB transformations to render $m_\ell$ and $M_R$
diagonal, while $m_D$ has the form of Eq.~(\ref{mDtri}). A specific class of
models which satisfy the above necessary condition in a trivial way are those
for which $U=\openone$, leading to $m_D=v\,Y_\triangle$. This condition is
necessary but not sufficient to allow for a prediction of the sign of the CP
asymmetry in neutrino oscillations, given the observed sign of the BAU together
with the low-energy data. Therefore, we will consider next a more restrictive
class of matrices $m_D$ of this form and we will show that, in an appropriate
limiting case, our structures for $m_D$ lead to the ones assumed by Frampton,
Glashow and Yanagida in \cite{Frampton:2002qc}.

\section{Minimal Scenarios}
\label{minimal}

From the analysis carried out in the previous section, it becomes clear that
the computation of the cosmological baryon asymmetry $Y_B$ in leptogenesis
scenarios strongly depends on the Yukawa structure of the Dirac neutrino mass
term and on the heavy Majorana neutrino mass spectrum. Moreover, if one assumes
that the seesaw mechanism is responsible for the smallness of the neutrino
masses, then the connection between the baryon asymmetry and low-energy
neutrino physics is unavoidable. In fact, this constitutes an advantage for the
leptogenesis mechanism when compared to other baryogenesis scenarios. In the
context of supersymmetric extensions of the SM it is possible (although not
always simple) to combine the study of leptogenesis and neutrino physics with
other physical phenomena like flavor-violating decays
\cite{Ellis:2002eh,Ellis:2002xg}. In general, this analysis does not give us
definite answers, yet it may help to discriminate among certain neutrino mass
models.

Recently, a considerable amount of work has been done aiming at relating viable
leptogenesis to all the available low-energy neutrino data coming from solar,
atmospheric and reactor experiments \cite{Buchmuller:2001dc}. Roughly speaking,
two different approaches to the problem are to be found. The first one is based
on the computation of the baryon asymmetry as a function of the lightest heavy
Majorana neutrino mass $M_1$, the CP asymmetry $\varepsilon_1$ and the
so-called effective neutrino mass
$\tilde{m}_1=(m_D^{\dagger}\,m_D^{})_{11}/M_1$ \cite{Plumacher:1996kc}. By
solving the Boltzmann equations, this kind of analysis provides valuable
information on the ranges of these parameters that lead to an acceptable value
of $Y_B$. The weak point of this procedure lies on the fact that the input
information depends on quantities which are not sensitive to the full structure
of $Y_\nu$ and $M_R$ and, therefore, no further conclusions can be drawn about
the class of models which can lead to acceptable values of the input parameters
referred above. In fact, the values of $M_1$, $\tilde{m}_1$ and $\varepsilon_1$
should not be taken as independent parameters. The second approach is based
upon initial assumptions on the structure of $Y_\nu$ and $M_R$ at high energies
which are fixed by recurring to theoretical arguments like for example grand
unified theories or flavor symmetries. Although in this framework some
generality is lost, it has the advantage that one can compute the generated
baryon asymmetry and, simultaneously, perform a low-energy neutrino data
analysis. It is precisely the latter approach that we shall follow in the
present work.

In this section we present a class of minimal scenarios for leptogenesis based
on the triangular decomposition of $Y_\nu$ given in Eqs.~(\ref{mDtri}) and
(\ref{Ytri1}). Namely, we would like to answer the question of how simple can
the structure of the Dirac neutrino Yukawa coupling matrix be in order not only
to get an acceptable value of the baryon asymmetry but also to accommodate the
neutrino data provided by the atmospheric, solar and reactor neutrino
experiments. In particular, we require the non-vanishing of the CP asymmetry
generated in the decays of the lightest heavy Majorana neutrino, since the
final value of the baryon asymmetry crucially depends on this quantity.
Throughout our analysis we shall also consider the predictions on the
CP-violating effects at low energies.

In the previous section we have seen that the Dirac neutrino Yukawa coupling
matrix $Y_\nu$ can be decomposed into the product of a unitary matrix $U$ and a
lower-triangular matrix $Y_\triangle$ (cf. Eq.~\ref{mDtri}). It was also shown
that the CP asymmetries $\varepsilon_j$ generated in the heavy Majorana
neutrino decays do not depend on the matrix $U$. In the special case
$U=\openone$ it is possible to establish the connection between leptogenesis,
low-energy CP violation and neutrino mixing, since the same phases affect these
phenomena. We classify this scenario as a minimal scenario for leptogenesis and
CP violation in the sense that the CP-violating sources that do not contribute
to leptogenesis are neglected. On the other hand, if $U \neq \openone$ this
connection is not trivial. Therefore, from now on we will consider the case
$U=\openone$ which implies the following simple structure for the Dirac
neutrino mass matrix:
\begin{align}
m_D=v\,Y_\triangle = v\,\left(\begin{array}{ccc}
 y_{11}   & \; 0       &\; 0 \\
 y_{21}\,e^{i\,\phi_{21}}   &\; y_{22}       &\; 0 \\
y_{31}\,e^{i\,\phi_{31}}   &\; y_{32}\,e^{i\,\phi_{32}} &\; y_{33}
\end{array}\right)\label{MDmin}\,.
\end{align}
Then, from Eq.~(\ref{epsj1}) the CP asymmetry generated in the decay of the
heavy Majorana neutrino $N_j$ is
\begin{align}
\label{ejtri} \varepsilon_j=-\frac{1}{8\pi(H_\triangle)_{jj}} \sum_{i\neq
j}\,\im[(H_\triangle)_{ij}^2]\,f_{ij}\,,
\end{align}
where
\begin{align}
\label{fij}
 H_\triangle =Y^{\dag}_\triangle Y_\triangle^{} \quad,\quad
 f_{ij}=f\!\left(\frac{M_i^{\,2}}{M_j^{\,2}}\right)\,,
\end{align}
with $f(x)$ defined in Eq.~(\ref{f}).

From Eqs.~(\ref{MDmin}) and (\ref{fij}) we readily obtain
\begin{align} &\im[(H_\triangle)_{21}^2]=y_{21}^2\,
   y_{22}^2\sin
(2{{\phi}_{21}}) + 2\,y_{21}\,y_{22}\,y_{31}\,y_{32}\sin
\theta_1\,
    + y_{31}^2\,y_{32}^2\sin \theta_2\,,\nl
&\im[(H_\triangle)_{31}^2]=y_{31}^2\,y_{33}^2\,\sin (2\,{{\phi
}_{31}})\,,\nl %\quad,\quad
&\im[(H_\triangle)_{32}^2]=\,y_{32}^2\,y_{33}^2\sin (2\,{{\phi
}_{32}})\,,\label{imhij}
\end{align} with
$\theta_1={{\phi }_{21}} + {{\phi }_{31}} - {{\phi }_{32}}$ and
$\theta_2=2\,\left( {{\phi }_{31}} - {{\phi }_{32}} \right)$.

All the information about neutrino masses and mixing is fully contained in the
effective neutrino mass matrix $\mathcal{M}_\nu$ which is determined through
the seesaw formula given by Eq.~(\ref{ssaw}). In this case
\begin{align}
\label{Mntri}
\mathcal{M}_\nu=\frac{v^2}{M_1}\left(\begin{array}{ccc} y_{11}^2
&\;y_{11}y_{21}e^{i\,\phi_{21}}\; &\;y_{11}
y_{31}e^{i\,\phi_{31}}\;\\ y_{11}y_{21}e^{i\,\phi_{21}} &y_{21}^2
e^{2i\,\phi_{21}}+y_{22}^2\frac{M_1}{M_2} &y_{21}
y_{31}e^{i(\phi_{31}+\phi_{21})}+y_{22}y_{32}
\frac{M_1}{M_2}\,e^{i\phi_{32}} \\ y_{11} y_{31}e^{i\phi_{31}}
&y_{21}
y_{31}e^{i(\phi_{31}+\phi_{21})}+y_{22}y_{32}\frac{M_1}{M_2}\,
e^{i\phi_{32}}
&y_{31}^2e^{2i\phi_{31}}+y_{33}^2\frac{M_1}{M_3}+y_{32}^2\frac{M_1}{M_2}
\,e^{2i\phi_{32}}
\end{array}\right)\,.
\end{align}
It follows from Eqs.~(\ref{ejtri})-(\ref{Mntri}) that, in principle, one can
obtain simultaneously viable values for the CP asymmetries $\varepsilon_j$ and
a phenomenologically acceptable effective neutrino mass matrix in order to
reproduce the solar, atmospheric and reactor neutrino data. This can be
achieved by consistently choosing the values of the free parameters $y_{ij}$,
$M_i$ and $\phi_{ij}\ $. Yet, a closer look at Eqs.~(\ref{ejtri})-(\ref{imhij})
shows that there are terms contributing to $\varepsilon_j$ which vanish
independently from the others. This means that a non-vanishing value of
$\varepsilon_j$ can be guaranteed even for simpler structures for $Y_\nu$,
which can be obtained from $Y_\triangle$ assuming additional zero entries in
the lower triangle\footnote{Notice however that the vanishing of diagonal
elements in $Y_\triangle$ would imply ${\rm det}\,(m_D)=0$ and consequently,
${\rm det}\,(\mathcal{M}_\nu)=0$, leading to the existence of massless light
neutrinos.}. The results are given in Table~\ref{table1} where we present the
textures constructed from $Y_\triangle$ by neglecting one (textures I-III) and
two (textures IV-VI) off-diagonal entries. The form of the effective neutrino
mass matrix $\mathcal{M}_\nu$ and the expressions for the CP asymmetries
$\varepsilon_{1,2}$ for each case are also given\footnote{In commonly used
language, textures I-III and IV-VI belong to the classes of four and five
texture zero matrices, respectively. For a complete discussion on seesaw
realization of texture-zero mass matrices see e.g. \cite{Kageyama:2002zw} and
for its implications at low energies see e.g. \cite{Frampton:2002yf}.}.

\begin{turnpage}
\squeezetable
\begin{table*}
\caption{Minimal textures based on the triangular decomposition of $m_D$ and
their respective light neutrino mass matrix $\mathcal{M_\nu}$ and
CP-asymmetries $\varepsilon_1$ and $\varepsilon_2$. } \label{table1}
\medskip
\newcommand{\cc}[1]{\multicolumn{1}{c}{#1}}
\begin{tabular}{clll}
\hline\noalign{\medskip} \normalsize{Texture} &\cc{
\normalsize{$Y_{\triangle}$}}
&\cc{\normalsize{$\frac{M_1}{v^2}\mathcal{M_\nu}$}}
& \cc{\normalsize{$\varepsilon_{1,2}$}}  \\
\noalign{\medskip}\hline
\noalign{\medskip} \normalsize{I}
&$\left(\begin{array}{ccc}
 y_{11}   & 0       &0 \\
 y_{21}\,e^{i\,\phi_{21}}  &y_{22}       &0 \\
0   & y_{32}\,e^{i\,\phi_{32}} &y_{33}
\end{array}\right)$ &$\left(\begin{array}{ccc}
y_{11}^2                             &y_{11}\,y_{21}\,e^{i\,\phi_{21}}       &0 \\
y_{11}\,y_{21}\,e^{i\,\phi_{21}}     & y_{21}^2\,e^{2i
{{\phi}_{21}}} +
  \frac{M_1}{M_2}\,y_{22}^2      &y_{22}\,y_{32}\,\frac{M_1}{M_2}\,e^{i\,\phi_{32}} \\
0   & y_{22}\,y_{32}\,\frac{M_1}{M_2}\,e^{i\,\phi_{32}}
&y_{33}^2\,\frac{M_1}{M_3}+y_{32}^2\,\frac{M_1}{M_2}\,e^{2\,i\,\phi_{32}}
\end{array}\right)$ &$\begin{array}{l} \varepsilon_1^{\rm \, I}
=-\dfrac{y_{21}^2\,y_{22}^2\,\sin(2\,\phi_{21})}{8\pi\,(\,y_{11}^2+y_{21}^2\,)}\,
f_{21} \\ \\ \varepsilon_2^{\rm
\,I}=\dfrac{y_{21}^2\,y_{22}^2\,\sin(2\,\phi_{21})\,
f_{12}-y_{32}^2\,y_{33}^2\,\sin(2\,\phi_{32})\,f_{32}}
{8\pi\,(\,y_{22}^2+y_{32}^2\,)}\,
 \end{array}$\\
\noalign{\bigskip}
\normalsize{II} &$\left(\begin{array}{ccc}
 y_{11}   & 0       &0 \\
 0  &y_{22}       &0 \\
 y_{31}\,e^{i\,\phi_{31}}   & y_{32}\,e^{i\,\phi_{32}} &y_{33}
\end{array}\right)$
&$\left(\begin{array}{ccc}
y_{11}^2   &0       &y_{11}\,y_{31}\,e^{i\,\phi_{31}} \\
0   &  y_{22}^2\,\frac{M_1}{M_2}      &y_{32}\,y_{22}\,\,
\frac{M_1}{M_2}e^{i\,\phi_{32}} \\
y_{11}\,y_{31}\,e^{i\,\phi_{31}}   &
y_{32}\,y_{22}\,\,\frac{M_1}{M_2}e^{i\,\phi_{32}}
&y_{31}^2\,e^{2\,i\,\phi_{31}}+y_{33}^2\,\frac{M_1}{M_3}+y_{32}^2\,
\frac{M_1}{M_2}\,e^{2\,i\,\phi_{32}}
\end{array}\right)$
&$\begin{array}{l} \varepsilon_1^{\rm \,II} =\dfrac{y_{31}^2
y_{32}^2\sin[2(\phi_{32}-\phi_{31})]\,f_{21}- y_{31}^2
y_{33}^2\sin(2\phi_{31})f_{31}}{8\pi(y_{11}^2+y_{31}^2)}
\\ \\ \varepsilon_2^{\rm
\,II}=\dfrac{y_{31}^2
y_{32}^2\sin[2(\phi_{31}-\phi_{32})]\,f_{12}- y_{32}^2
y_{33}^2\sin(2\phi_{32})f_{32}}{8\pi(y_{22}^2+y_{32}^2)}
\end{array}$
\\
\noalign{\bigskip} \normalsize{III} &$\left(\begin{array}{ccc}
 y_{11}   & 0       &0 \\
 y_{21}\,e^{i\,\phi_{21}}  &y_{22}       &0 \\
y_{31}\,e^{i\,\phi_{31}}   & 0 &y_{33}
\end{array}\right)$
&$\left(\begin{array}{ccc}
y_{11}^2   &y_{11}\,y_{21}\,e^{i\,\phi_{21}}       &y_{11}\,y_{31}\,e^{i\,\phi_{31}} \\
y_{11}\,y_{21}\,e^{i\,\phi_{21}}   & y_{21}^2\,e^{2i\phi_{21}}+
y_{22}^2\,\frac{M_1}{M_2}
& y_{21}\,y_{31}\,e^{i(\phi_{21}+\phi_{31})}   \\
y_{11}\,y_{31}\,e^{i\,\phi_{31}}
&y_{21}\,y_{31}\,e^{i(\phi_{21}+\phi_{31})}
&y_{31}^2\,e^{2\,i\,\phi{31}}+y_{33}^2\,\frac{M_1}{M_3}
\end{array}\right)$
&$\begin{array}{l} \varepsilon_1^{\rm \,III} =-\dfrac{y_{21}^2
y_{22}^2\sin(2\phi_{21})f_{21}+ y_{31}^2
y_{33}^2\sin(2\phi_{31})f_{31}}{8\pi(y_{11}^2+y_{21}^2+y_{31}^2)}
\\ \\ \varepsilon_2^{\rm \,III}=\dfrac{y_{21}^2
y_{22}^2\sin(2\phi_{21})}{8\pi\, y_{22}^2}f_{12}\end{array}$
\\
\noalign{\bigskip}
\normalsize{IV} &$ \left(\begin{array}{ccc}
y_{11}   &0       & 0 \\ y_{21}\,e^{i\,\phi_{21}}   &y_{22}       &0 \\
0   &0 &y_{33} \end{array}\right)$ &$\left(\begin{array}{ccc}
y_{11}^2   &y_{11}\,y_{21}\,e^{i\,\phi_{21}}       & 0 \\
y_{11}\,y_{21}\,e^{i\,\phi_{21}}   & y_{21}^2\,e^{2\,i\,\phi_{21}}
+y_{22}^2\,\frac{M_1}{M_2}      & 0   \\
0   &0 &y_{33}^2\,\frac{M_1}{M_3}
\end{array}\right)$ &$\begin{array}{l} \varepsilon_1^{\rm \, IV}
=-\dfrac{y_{21}^2\,y_{22}^2\,\sin(2\,\phi_{21})}{8\pi\,(\,y_{11}^2+y_{21}^2\,)}\,
f_{21} \\ \\\varepsilon_2^{\rm \, IV}
=\dfrac{y_{21}^2\,y_{22}^2\,\sin(2\,\phi_{21})}{8\pi\,y_{22}^2}\,
f_{12} \end{array}$\\
\noalign{\bigskip}
\normalsize{V} &$\left(\begin{array}{ccc}
 y_{11}   & 0       &0 \\
0   &y_{22}       &0 \\
 y_{31}\,e^{i\,\phi_{31}}   &0&y_{33}
\end{array}\right)$
&$\left(\begin{array}{ccc}
y_{11}^2   &0       &y_{11}\,y_{31}\,e^{i\,\phi_{31}} \\
0   &  y_{22}^2\,\frac{M_1}{M_2}      & 0   \\
y_{11}\,y_{31}\,e^{i\,\phi_{31}}   &0
&y_{31}^2\,e^{2\,i\,\phi_{31}}+y_{33}^2\,\frac{M_1}{M_3}
\end{array}\right)$
&$\begin{array}{l} \varepsilon_1^{\rm \, V}
=-\dfrac{y_{31}^2\,y_{33}^2\,\sin(2\,\phi_{31})}{8\pi\,(\,y_{11}^2+y_{31}^2\,)}\,
f_{31} \\ \\ \varepsilon_2^{\rm \,V}=0\end{array}$
\\
\noalign{\bigskip}
\normalsize{VI} &$ \left(\begin{array}{ccc}
 y_{11}   & 0       &0 \\
 0  &y_{22}       &0 \\
0   & y_{32}\,e^{i\,\phi_{32}} &y_{33}
\end{array}\right)$
&$\left(\begin{array}{ccc}
y_{11}^2   &0       &0 \\
0   &  y_{22}^2\,\frac{M_1}{M_2}      &y_{22}\,y_{32}\,
\frac{M_1}{M_2}\,e^{i\,\phi_{32}} \\
0   & y_{22}\,y_{32}\,\frac{M_1}{M_2}\,e^{i\,\phi_{32}}
&y_{33}^2\,\frac{M_1}{M_3}+y_{32}^2\,\frac{M_1}{M_2}\,
e^{2\,i\,\phi_{32}}
\end{array}\right)$
&$\begin{array}{l} \varepsilon_1^{\rm \, VI} =0
\\ \\ \varepsilon_2^{\rm
\,VI}=-\dfrac{y_{32}^2\,y_{33}^2\,\sin(2\,\phi_{32})}{8\pi\,
(\,y_{22}^2+y_{32}^2\,)}f_{32}
\end{array}$\\\noalign{\bigskip}\hline
\end{tabular}
\end{table*}
\end{turnpage}

Let us first discuss textures IV-VI. For these three textures the effective
neutrino mass matrix $\mathcal{M}_\nu$ predicts a complete decoupling of one
light neutrino from the other two. This is in disagreement with the available
neutrino data which indicates that, in the framework of the LMA solution, only
one neutrino mixing angle should be small, namely $\theta_{13}$, instead of two
as predicted by textures IV-VI. Furthermore, texture VI predicts a vanishing
value for the CP asymmetry in the decay of the lightest heavy Majorana
neutrino, implying $Y_B=0$ in the case of hierarchical heavy Majorana
neutrinos. Texture III can also be excluded on the grounds that it cannot
account for the large solar angle. To illustrate this, let us write for this
texture the matrix $L$ given in Eq.~(\ref{MnuLL}) as
\begin{equation}
L = i\,m_D\,D_R^{-1/2} \equiv\left(\begin{array}{ccc}
z_1 &\;\; 0 &\;\;0 \\ z_2 &\;\;y_2 &\;\;0 \\
z_3 &\;\; 0 &\;\;x_3 \end{array}\right)\,,
\end{equation}
where $z_1$, $y_2$ and $x_3$ are real and positive \cite{Lavignac:2002gf}. We
also take $z_2$ and $z_3$ real to simplify the discussion. Considering first
the hierarchical case $\dmatm \simeq  m_3^2\ $, a large atmospheric angle and
$m_2 < m_3$ naturally arise if $z_2 \sim z_3 > y_2,x_3\ $. More precisely,
$\tan \theta_{23} \simeq z_2/z_3\ $. In addition, this implies
\begin{equation}
\tan \theta_{13} \simeq \frac{z_1}{\sqrt{z_2^2+z_3^2}} \equiv
t_{13}\,,
\end{equation}
which has to satisfy the CHOOZ bound, $t_{13} \le 0.2$ \cite{Apollonio:1999ae}.
The solar angle is approximately given by
\begin{equation}
\tan( 2 \theta_{12}) \simeq 2\,t_{13}\,\frac{y_2^2+x_3^2}{
\frac{z_3}{z_2}\,y_2^2 + \frac{z_2}{z_3}\,x_3^2 - t_{13}
(\frac{z_2}{z_3}\,y_2^2 + \frac{z_3}{z_2}\,x_3^2) } \le
\frac{2\,t_{13}\,\beta }{1-t_{13}\,\beta^2} \equiv B_{12}(\beta,
t_{13})\,,
\end{equation}
where $\beta \equiv {\rm max}(\tan \theta_{23}, \cot \theta_{23})$. The upper
bound $B_{12}(\beta, t_{13}=0.2)$ is an increasing function of $\beta$.
Experimentally  $0.7 \le \beta \le 1.7$, so that $B_{12}^{\rm max} =
B_{12}(1.7,0.2) = 4$. This corresponds to $\tan^2\theta_{12} \le 0.3$, while
the fitted LMA solution requires $\tan^2\theta_{12} > 0.3$. We thus conclude
that type III cannot account for the observed large solar angle in the
hierarchical case. Moreover, it turns out that it cannot also account for the
pattern of ${\cal M}_\nu$ required by inverse hierarchy since, in this case,
the atmospheric oscillation fit would require $z_1 z_2 \simeq z_1 z_3 \simeq
\sqrt{\dmatm}$ and $z_1^2, z_2^2, z_3^2 \ll \sqrt{\dmatm}$. Finally, the
degenerate case can be accommodated only by tuning the elements of $L$. In view
of this, from now on we will focus our analysis only on textures I and II.

Another interesting fact which comes out from the observation of
Table~\ref{table1} is that the phase content for textures I-II can be further
reduced. Indeed, it is straightforward to show that with only one phase in
$Y_\triangle$ it is possible to obtain a non-vanishing CP asymmetry
$\varepsilon_1$.

As it has been stated in Section~\ref{generframe}, the strength of CP violation
at low energies is sensitive to the value of the CP invariant
$\mathcal{J}_{CP}$ given by Eq.~(\ref{Jfin}). For the textures I and II we get:
\begin{align}
\label{JI}  \mathcal{J}_{CP}^{\rm \, I}=&\frac{y_{11}^2\,
y_{21}^2\, y_{32}^2\,y_{22}^2\,v^{12}}{M_1^3 M_2^3\dmtwon \dmthon
\dmthtw}\left[(y_{21}^2\, y_{32}^2 +y_{11}^2\,y_{22}^2+y_{11}^2\,
y_{32}^2)\sin(2\phi_{21})-y_{22}^2\,y_{33}^2
\frac{M_1}{M_3}\sin(2\phi_{32})\right.\nl & +\left.y_{33}^2
(y_{11}^2+ y_{21}^2)\frac{M_2}{M_3}\sin\left[2\left(\phi_{21} -
\phi_{32} \right)\right] \right]  \,,\nonumber \\
 \mathcal{J}_{CP}^{\rm \, II}=&\frac{y_{11}^2\, y_{22}^2\,
y_{31}^2\, y_{32}^2\,v^{12}}{M_1^3 M_2^3\dmtwon \dmthon
\dmthtw}\left[(y_{22}^2\,y_{31}^2+y_{11}^2\,y_{22}^2+y_{11}^2\,
y_{32}^2)\,\sin\left[2\,(\phi_{32} -\phi_{31})\right]\right.\nl &
\left.+\,y_{22}^2\,y_{33}^2
\frac{M_1}{M_3}\,\sin(2\phi_{32})-y_{11}^2\,y_{33}^2\,
\frac{M_2}{M_3}\,\sin(2\phi_{31})
 \right]\,.
\end{align}%
From Table~\ref{table1} and Eqs.~(\ref{JI}) it is clear that we can have both
$\varepsilon_1 \neq 0$ and $\mathcal{J}_{CP} \neq 0$ with a single
non-vanishing phase.

 \bigskip \medskip
{\bf On the relative sign between \bm{$Y_B$} and \bm{$\mathcal{J
}_{CP}$}} %\label{sign}
\bigskip \medskip

It has recently been pointed out in \cite{Frampton:2002qc} that the relative
sign between CP violation and the baryon asymmetry can be predicted in a
specific class of models. In our framework, it is clear from Table~\ref{table1}
and Eqs.~(\ref{JI}) that the relative sign between the low-energy invariant
$\mathcal{J}_{CP}$ and the CP asymmetries $\varepsilon_j$ cannot be predicted
without specifying the values of the heavy neutrino masses $M_i$ and the Dirac
Yukawa couplings\footnote{For the sake of more generality, we do not assume the
mass-ordering $M_1 < M_2$ in this discussion.}. This is mainly due to the fact
that at least one of these quantities depends on both phases appearing in the
Dirac neutrino Yukawa coupling matrix for textures I and II. Therefore, in
order to establish a direct connection between the sign of the CP asymmetries
$\varepsilon_j$ and the low-energy CP invariant $\mathcal{J}_{CP}$ further
assumptions are required. For instance, considering the case $\phi_{32}=0$ and
$M_1,M_2 \ll M_3$ so that the terms proportional to $f_{31}$ can be safely
neglected in $\varepsilon_{1}^{\rm II}$ (see Table \ref{table1}), we obtain
from Eqs.~(\ref{JI}) and Table \ref{table1} the relative signs given in Table
\ref{table2}.
\begin{table*}
\caption{Relative sign between the CP asymmetries $\varepsilon_{1,2}$ and the
low-energy CP invariant $\mathcal{J}_{CP}$ for textures I and II considering
the different heavy Majorana mass regimes. We assume $M_1,M_2 \ll M_3$ and
$\phi_{32}=0$. The parameter $x \simeq 0.4$ corresponds to the zero of the loop
function $f(x)$ defined in Eq.~(\ref{f}).}\label{table2}
\medskip
\newcommand{\cc}[1]{\multicolumn{1}{c}{#1}}
\renewcommand{\tabcolsep}{0.7pc}
\begin{tabular}{lcccc}
\hline\noalign{\smallskip}&$\dfrac{M_1}{M_2}< x$
&$x<\dfrac{M_1}{M_2}<1$ &$\dfrac{M_2}{M_1}< x$ &$x<\dfrac{M_2}{M_1}<1$ \\
\noalign{\smallskip} \hline
%Texture I   &           &          &          & \\
sgn$\,(\varepsilon_1^{\rm I}\cdot\mathcal{J}_{CP}^{\rm \, I})$ &$+$ &$+$ &$+$ &$-$\\
sgn$\,(\varepsilon_2^{\rm I}\cdot\mathcal{J}_{CP}^{\rm \, I})$   &$-$   &$+$ &$-$  &$-$ \\
\hline\noalign{\smallskip}
sgn$\,(\varepsilon_1^{\rm II}\cdot\mathcal{J}_{CP}^{\rm \, II})$   &$-$    &$-$  &$-$ &$+$ \\
sgn$\,(\varepsilon_2^{\rm II}\cdot\mathcal{J}_{CP}^{\rm \, II})$
&$+$ &$-$  &$+$ &$+$ \\ \hline\noalign{\smallskip}
\end{tabular} \end{table*}

It is well known that in the case of hierarchical heavy neutrinos, the sign of
$Y_B$ is fixed by the sign of the CP asymmetry generated in the decay of the
lightest heavy neutrino, let us say $\varepsilon_L$ . For $Y_B$ to be positive,
as required by the observations, one must have $\varepsilon_L < 0$. Thus, the
measurement of the sign of CP violation at low-energy could in principle
exclude some of the heavy Majorana neutrino mass regimes presented in
Table~\ref{table2}. As an example, let us assume that $\mathcal{J}_{CP}< 0$. In
this case, the requirement $\varepsilon_L < 0$ would exclude the mass regime
$M_2/M_1 < x$ for texture I and $M_1/M_2 < x$ for texture II.

Finally, it is interesting to note that the Dirac neutrino Yukawa matrices
considered in \cite{Frampton:2002qc} correspond to our textures I and II with
$y_{33}=0$ and have the remarkable feature that the number of real parameters
equals the number of masses and mixing angles. In this limit the heavy Majorana
neutrino $N_3$ completely decouples, rendering this situation
phenomenologically equivalent to the two heavy neutrino case considered in
\cite{Frampton:2002qc}. Namely, it can be easily seen that for $y_{33}=0$ the
phase $\phi_{32}$ can be rephased away and consequently the CP asymmetries
$\varepsilon_{1,2}$ and the low-energy CP invariant $\mathcal{J}_{CP}$ will
only depend on the phases $\phi_{21}$ and $\phi_{31}$ for textures I and II,
respectively. This means that the examples considered in \cite{Frampton:2002qc}
are special cases of our framework which is motivated by the condition that
$m_D$ does not contain any phase that would only contribute to low-energy CP
violation. It is clear from Table~\ref{table2} that, in general, even by
keeping only one phase the sign of $\varepsilon_j\cdot\mathcal{J}_{CP}$ depends
on the particular Yukawa texture and hierarchy between the heavy Majorana
neutrinos.
\section{Examples}
\label{example}

In this section we present some examples of the minimal textures discussed in
Section~\ref{minimal} and proceed to the study of their implications to
low-energy physics as well as to the computation of the baryon asymmetry $Y_B$
through the numerical solution of the Boltzmann equations as described in the
Appendix. We will only consider cases that lead to the LMA solution of the
solar neutrino problem, which means that the neutrino mixing angles and the
squared mass differences lie in the typical ranges:
\begin{align}
\label{LMAdt} 2.5 \times 10^{-5}\,\text{eV}^2 & \lesssim \dmsol
\lesssim 3.4 \times 10^{-4}\,\text{eV}^2 \quad,\quad 0.24 \lesssim
\tgsol \lesssim
0.89\,,\nonumber\\
1.4 \times 10^{-3}\,\text{eV}^2 & \lesssim \dmatm \lesssim 6.0 \times
10^{-3}\,\text{eV}^2 \quad,\quad 0.40 \lesssim \tgatm \lesssim 3.0\,,
\end{align}
with the solar and atmospheric mixing angles being identified as $
\theta_\odot \equiv \theta_{12} $, $ \theta_ @ \equiv \theta_{23} $,
respectively, in the standard parametrization \cite{Hagiwara:pw}. The
$\theta_{13}$ mixing angle is at present constrained by reactor neutrino
experiments to satisfy $|\sin \theta_{13}| \equiv |U_{e3}| \lesssim 0.2$
\cite{Apollonio:1999ae}.

\subsection{Hierarchical heavy Majorana neutrinos $\bm{(M_1 < M_2 \ll M_3)}$ }
\label{hierarchical}%

The first example is a realization of the texture~I given in Table~\ref{table1}
with $\phi_{32}=0$. The entries of the Dirac neutrino Yukawa coupling and the
right-handed neutrino mass matrices are chosen to be of order:
\begin{align}
\label{YDex1} Y_\nu = \frac{\Lambda_D}{v}\left(\begin{array}{ccc}
 \epsilon^6   & \;\;0       &\;\; 0 \\
 \epsilon^6\,e^{i\,\phi_{21}}   &\;\;\epsilon^4 &\;\; 0 \\
 0   &\;\; \epsilon^4       &\;\; 1
\end{array}\right)\;\;,\;\;D_R=\Lambda_R\,{\rm
diag}\,(\epsilon^{12},\epsilon^{10},1)\,,
\end{align}
where $\epsilon < 1$ is a small parameter. For our numerical estimates we
consider the typical Dirac neutrino mass scale to be $\Lambda_D \simeq
100$~GeV, which corresponds approximately to the top quark mass at the GUT
scale \cite{Fusaoka:1998vc}. The neutrino mass matrix $\mathcal{M}_\nu$ is then
given by:
\begin{align}
\label{mnex1} \mathcal{M}_\nu=\frac{\Lambda_D^2}{\epsilon^2
\Lambda_R}\left(\begin{array}{ccc}
\epsilon^2                   &\quad\epsilon^2\,e^{i\,\phi_{21}}     &\quad0 \\
\epsilon^2\,e^{i\,\phi_{21}} &\quad 1+\epsilon^2\,e^{2i\,\phi_{21}} &\quad 1 \\
 0                           &\quad 1                    &\quad 1+\epsilon^2
\end{array}\right)\,.
\end{align}
 In this particular case, considering for the moment $\phi_{21}=0$,
one gets\footnote{We have checked that the light neutrino masses
$m_i$ and mixing angles $\theta_{ij}$ are not sensitive to the
phase $\phi_{21}$.}: \vspace{0.2cm}
\begin{align} \label{msex1} m_1\simeq
\frac{\Lambda_D^2}{2\Lambda_R}(2- \sqrt{2}) \quad,\quad m_2\simeq
\frac{\Lambda_D^2}{2\Lambda_R}(2+\sqrt{2}) \quad , \quad m_3\simeq
\frac{\Lambda_D^2}{\epsilon^2 \Lambda_R}(2+\epsilon^2)\,,
\end{align}
leading to
\begin{align}
\label{dmex1} \dmtwon \simeq
\frac{2\,\sqrt{2}\,\Lambda_D^4}{\Lambda_R^2}\quad,\quad \dmthtw \simeq
\frac{4\,\Lambda_D^4}{\epsilon^4\,\Lambda_R^2}\,.
\end{align}
The requirement
\begin{align}
\label{rsol} \left. \frac{\dmsol}{\dmatm} \right|_{\rm
LMA}\!\!\!\,=\,\frac{\dmtwon}
 {\dmthtw}\simeq \frac{\epsilon^4}{\sqrt{2}}\,,
\end{align}
forces $\epsilon$ to be in the range $0.3 \lesssim \epsilon
\lesssim 0.7$. For the neutrino mixing angles we have:
\begin{align}
\label{angl} \tan^2\theta_{12}\simeq 1-\frac{\epsilon^2}
{2\sqrt{2}} \quad,\quad \tan^2\theta_{23}\simeq 1 \quad,\quad
|U_{e3}| \simeq \frac{\epsilon^2}{2\,\sqrt{2}}\,,
\end{align}
which are in good agreement with the data taking into account the range of
$\epsilon$. The matrix $V_L$ which corresponds to the left-handed rotation
involved in the diagonalization of $m_D$ can be viewed as the equivalent in the
leptonic sector of the quark mixing matrix $V_{CKM}$. This matrix is obtained
by diagonalizing the Hermitian matrix $Y_\nu Y_\nu^{\,\dag}$,
\begin{align}
\label{VLex1} V_L \simeq \left(\begin{array}{ccc}
1-\epsilon^8/2              &\quad\epsilon^4-\epsilon^{12}/2 &\quad0 \\
-\epsilon^4+\epsilon^{12}/2   &\quad1-\epsilon^8/2           &\quad\epsilon^8 \\
 \epsilon^{12}                &\quad-\epsilon^8              &\quad 1
\end{array}\right)\,.
\end{align}
This is in fact an interesting result in the sense that we are in the presence
of a typical \emph{large mixing out of small mixing} situation, where the large
neutrino mixing is generated through the seesaw mechanism
\cite{Altarelli:1999dg,Akhmedov:2000yt}. The scale $\Lambda_R$, or equivalently
the mass $M_3$ of the heaviest Majorana neutrino, is determined by requiring
$\dmthtw$ to be in the experimental range given in (\ref{LMAdt}). We find
\begin{align}
\label{LR} M_3 = \Lambda_R \,\simeq \, \frac{2\,\Lambda_D^2}{\epsilon^2\,
\sqrt{\dmatm}}\, \simeq\,
\frac{2^{5/4}\,\Lambda_D^2}{\sqrt{\dmsol}}\,\simeq \, 8\times
10^{15}\,\,{\rm GeV}\,,
\end{align}
and Eq.~(\ref{YDex1}) implies $M_1 \simeq 1.3 \times 10^{11}$~GeV and $M_2
\simeq 8.4 \times 10^{12}$~GeV, for $\epsilon=0.4$.

From the expression of the CP asymmetry $\varepsilon_1^{\rm I}$ given in
Table~\ref{table1}, we obtain
\begin{align}
\label{e1ex1} \varepsilon_1
=\dfrac{3\,y_{21}^2\,y_{22}^2\,\sin(2\,\phi_{21})}{16\pi\,(\,
y_{11}^2+y_{21}^2\,)}\,\frac{M_1}{M_2} \simeq
\frac{3\,\Lambda_D^2\,\epsilon^{10}}{32\pi\,v^2} \sin(2\,
\phi_{21})\simeq 10^{-6}\sin(2\,\phi_{21})\,,
\end{align}
which is maximized for $\phi_{21}=\pi/4\,.$ The value of the CP invariant
$\mathcal{J}_{CP}$ can be computed from the expression
\begin{align}\label{Jex1} \mathcal{J}_{CP}=&\frac{y_{11}^2\, y_{21}^2
\, y_{32}^2\,y_{31}^2\,v^{12}}{M_1^3 M_2^3\dmtwon \dmthon
\dmthtw}\left[y_{21}^2\, y_{32}^2+y_{11}^2(\,y_{22}^2+
y_{32}^2)+y_{33}^2 (y_{11}^2+ y_{21}^2)\frac{M_2}{M_3}\right]
\sin(2\phi_{21}) \,,
 \end{align}
which is obtained from Eq.~(\ref{JI}) setting $\phi_{32}=0$. The result is
straightforward in this case
\begin{align}
\label{Jap1} \mathcal{J}_{CP} \simeq
\frac{3\,\epsilon^2\sin(2\phi_{21})}{32\,\sqrt{2}} \simeq 1.1
\times 10^{-2} \sin(2\,\phi_{21})\,,
\end{align}
rendering visible CP-violating effects in the next generation of neutrino
factories for $\phi_{21}\simeq \pi/4$. It is interesting to note from
Eqs.~(\ref{e1ex1}) and (\ref{Jap1}) that the dependence of $\varepsilon_1$ and
$\mathcal{J}_{CP}$ on the phase $\phi_{21}$ is such that both quantities are
simultaneously maximized. Notice also that $\varepsilon_1$ and
$\mathcal{J}_{CP}$ have the same sign, as it should be according to
Table~\ref{table2}.

\begin{figure*}
$$\includegraphics[width=12cm]{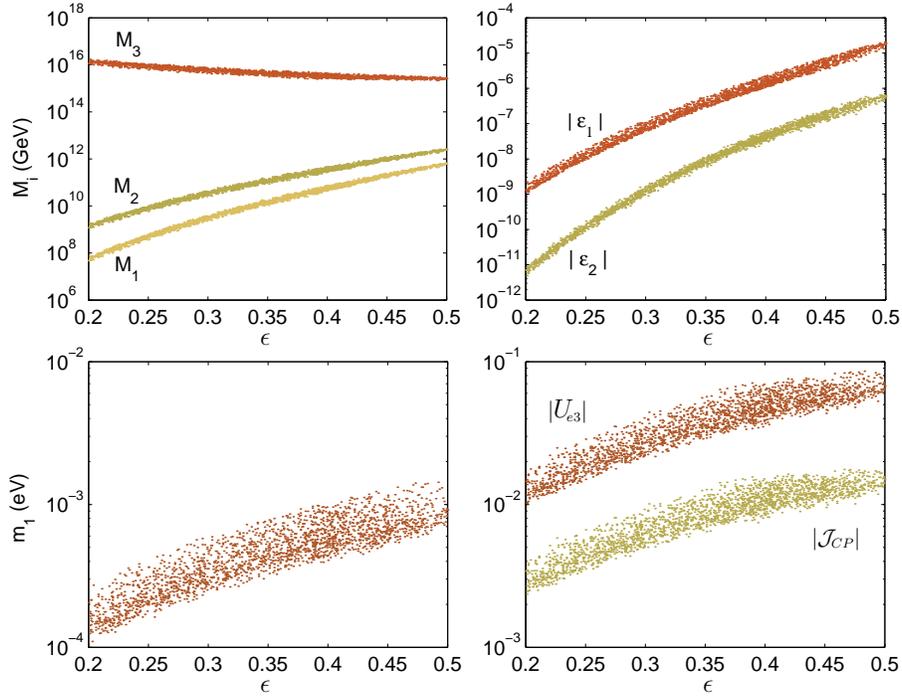}$$
\caption{Results of the numerical computation for the textures presented in
Eq.~(\ref{YDex1}) for $Y_\nu$ and $M_R$ with the CP-violating phase
$\phi_{21}=\pi/4$. The heavy Majorana neutrino masses $M_i$ and the CP
asymmetries $\varepsilon_1$, $\varepsilon_2$ are plotted as functions of the
texture parameter $\epsilon$. The dependence of $|U_{e3}|$ and
$\mathcal{J}_{CP}$ on $\epsilon$ is also shown. The dotted areas satisfy the
neutrino constraints (\ref{LMAdt}) for the LMA solar solution.} \label{fig1}
\end{figure*}

All the results presented above are confirmed by the full numerical computation
presented in Fig.~\ref{fig1}, where we have randomly included $\mathcal{O}$(1)
coefficients\footnote{For illustration, we have taken the $\mathcal{O}$(1)
coefficients in the range $[0.9,1.3]$. We notice however that the results are
not too sensitive to slight variations of this range.} for the non-vanishing
entries of $Y_\nu$ and taken $\phi_{21}=\pi/4$. The spreading of the points in
the figure, due to the random variation of the coefficients, shows that the
textures are
stable under these perturbations.\\
\begin{figure*}
$$\includegraphics[width=10cm]{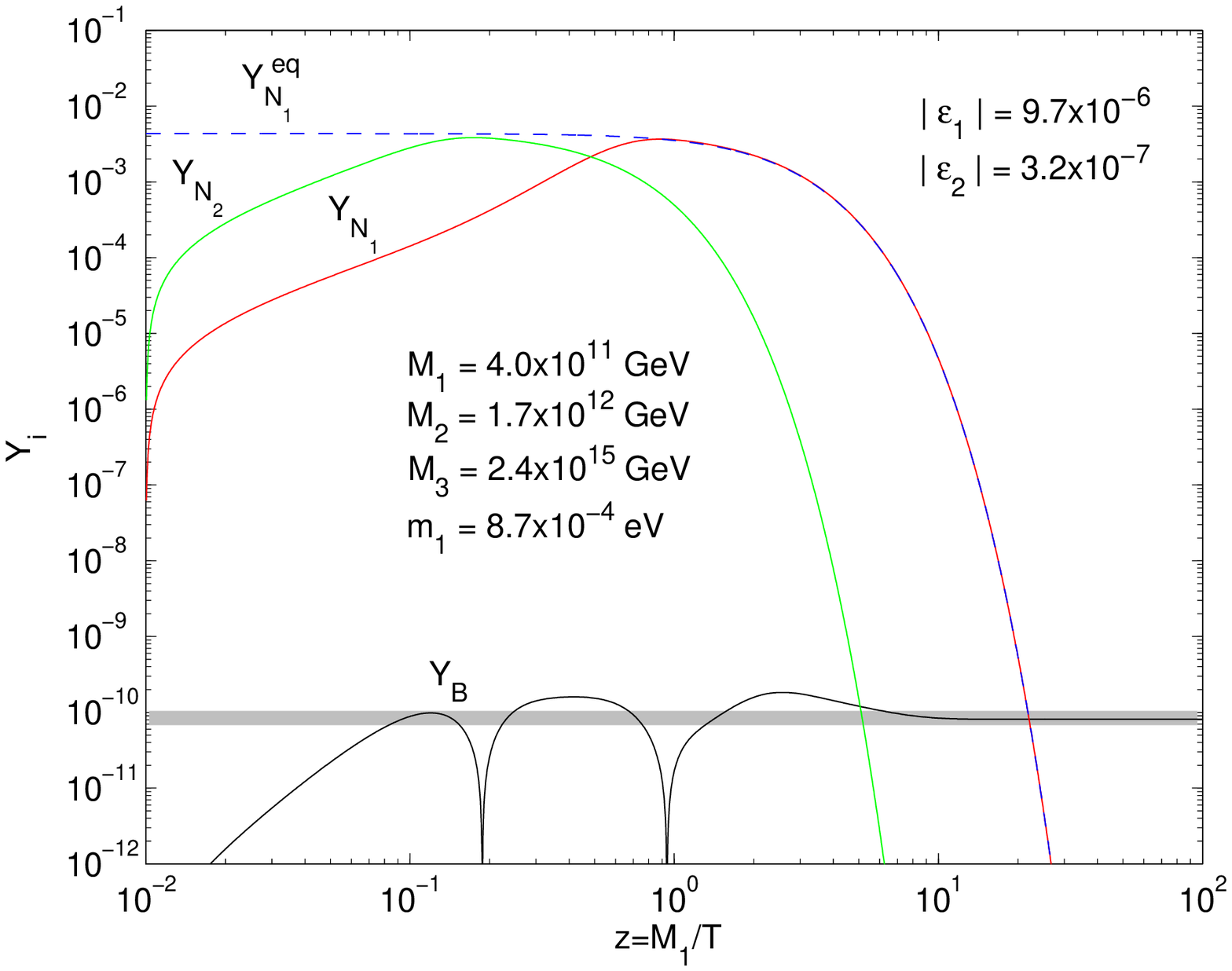}$$
\caption{Results for the distributions $Y_{N_1}$, $Y_{N_2}$ and for the baryon
asymmetry $Y_B$ as functions of $z=M_1/T$, obtained from the numerical solution
of the Boltzmann equations. The plot refers to the texture for $Y_\nu$
considered in Eq.~(\ref{YDex1}) with $\phi_{21}=\pi/4$ and $\epsilon=0.5$. The
values of the heavy Majorana neutrino masses $M_i$, the CP asymmetries
$\varepsilon_{1,2}$ and the lightest neutrino mass $m_1$ are consistent with
the ones presented in Fig.~\ref{fig1} for this value of $\epsilon$. The value
predicted for the final baryon asymmetry is $Y_B \simeq 9 \times 10^{-11}$.}
\label{fig2}
\end{figure*}
In order to compute the value of the baryon asymmetry we proceed to the
numerical solution of the Boltzmann equations as described in the Appendix,
taking the initial conditions: $Y^{\,0}_{N_i}=0 \,,\, Y_{B-L}^{\,0}=0$. The
results are presented in Fig.~\ref{fig2} where we have plotted $Y_{N_1}$,
$Y_{N_2}$ and $Y_B$ as functions of the parameter $z=M_1/T$ for given values of
the CP asymmetries, heavy Majorana neutrino masses and the lightest neutrino
mass. The predicted value for the final baryon asymmetry is $Y_B \simeq 9\times
10^{-11}$, which is inside the observational range (\ref{YBrng}).

Our next example is a particular case of the texture II presented in
Table~\ref{table1} with $\phi_{32}=0$. The Dirac neutrino Yukawa coupling and
heavy neutrino mass matrices are chosen in this case to be of order
\begin{align}\label{YDex2}
Y_\nu = \frac{\Lambda_D}{v}\left(\begin{array}{ccc}
 \epsilon^5                       &\quad 0            &\quad 0 \\
 0                                &\quad \epsilon^3   &\quad 0 \\
 \epsilon^5 e^{i\,\phi_{31}}   &\quad \epsilon^3      &\quad 1
\end{array}\right)\;\;,\;\;D_R=\Lambda_R\,{\rm
diag}\,(\epsilon^{10},\epsilon^{8},1)\,,
\end{align}
leading to the following light neutrino mass matrix
\begin{align}
\label{mnex2} \mathcal{M}_\nu=\frac{\Lambda_D^2}{\epsilon^2
\Lambda_R}\left(\begin{array}{ccc}
\epsilon^2                    &\quad 0 &\quad\epsilon^2\,e^{i\,\phi_{31}} \\
0                             &\quad 1 &\quad 1 \\
 \epsilon^2\,e^{i\,\phi_{31}} &\quad 1 &\quad 1+ \epsilon^2(1+e^{2\,i\,\phi_{31}})
\end{array}\right)\,.
\end{align}
The predictions for the $\Delta m_{ij}^2$'s and neutrino mixing angles are
similar to the ones given in Eqs.~(\ref{dmex1}) and (\ref{angl}). Moreover,
from Eqs.~(\ref{LR}) and (\ref{YDex2}) and for $\epsilon=0.4$, we get the
following heavy Majorana neutrino masses: $M_1 \simeq 9 \times 10^{11} \, {\rm
GeV} \,,\, M_2 \simeq 6 \times 10^{12}\, {\rm GeV} \,,\, M_3 \simeq 8 \times
10^{15}\, {\rm GeV}$. The left-handed matrix $V_L$ is given in this case by:
\begin{align}
\label{VLex2} V_L \simeq \left(\begin{array}{ccc}
-1               &\quad -\epsilon^{10}  &\quad \epsilon^{10} \\
-\epsilon^{10}   &\quad 1               &\quad\epsilon^6 \\
\epsilon^{10}    &\quad-\epsilon^6      &\quad 1
\end{array}\right)\,.
\end{align}

Finally, the CP asymmetry can be obtained from the expression of
$\varepsilon_1^{\rm \,II}$ presented in Table~\ref{table1} with $\phi_{32}=0$.
Taking into account the form of $Y_\nu$ and $D_R$ as in Eqs.~({\ref{YDex2}}),
we obtain
\begin{figure*}
$$\includegraphics[width=12cm]{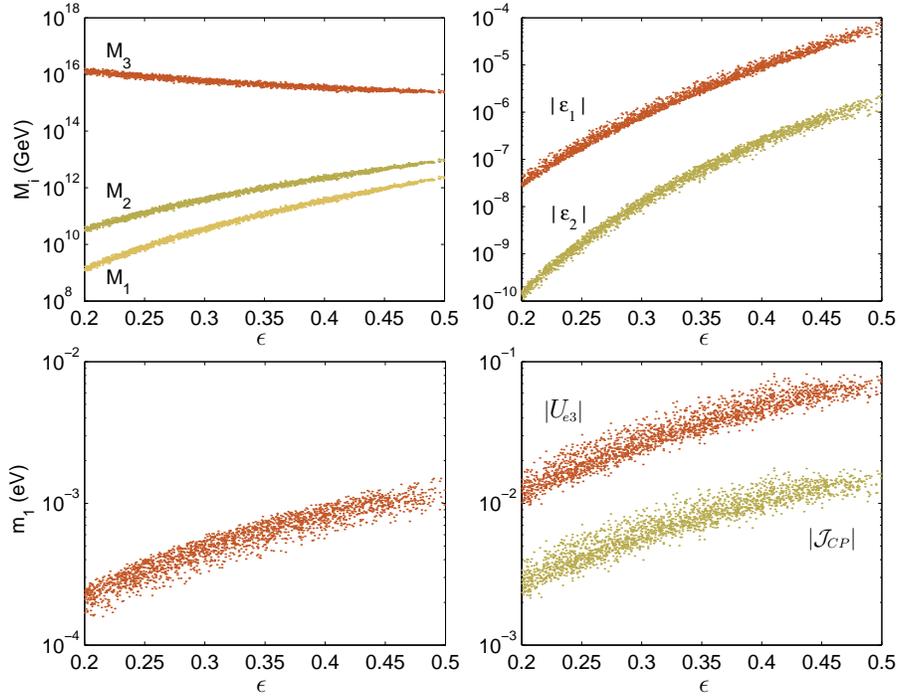}$$
\caption{The same as in Fig.~\ref{fig1} for the textures considered in
Eq.~(\ref{YDex2}) and $\phi_{31}=\pi/4$.} \label{fig3}
\end{figure*}
\begin{align}
\label{e1ex2} \varepsilon_1
=\dfrac{3\,y_{31}^2\,\sin(2\phi_{31})}{16\,\pi(y_{11}^2+y_{31}^2)}\left(
y_{32}^2\,\frac{M_1}{M_2}+y_{33}^2\,\frac{M_1}{M_3}\right) \simeq
\frac{3\,\Lambda_D^2\,\epsilon^8}{32\,\pi\,v^2}\,\sin(2\,\phi_{31})
\simeq 6.5 \times 10^{-6}\,\sin(2\,\phi_{31})\,,
\end{align}
where the last estimate has been obtained assuming $\epsilon=0.4$. It can also
be shown that the CP invariant $\mathcal{J}_{CP}$ reads (see Eq.~(\ref{JI})),
\begin{align}
\label{Jex2}
 \mathcal{J}_{CP}=-\frac{y_{11}^2\, y_{22}^2\,
y_{31}^2\, y_{32}^2\,v^{12}}{M_1^3 M_2^3\dmtwon \dmthon
\dmthtw}\left[y_{22}^2\,y_{31}^2+y_{11}^2\,(y_{22}^2+
y_{32}^2)+y_{11}^2\,y_{33}^2\,\frac{M_2}{M_3}
 \right]\,\sin(2\phi_{31})\,,
\end{align}
which leads to the approximate result
\begin{align}
\mathcal{J}_{CP} \simeq -
\frac{3\,\epsilon^2\sin(2\phi_{31})}{32\,\sqrt{2}} \simeq -1.1
\times 10^{-2} \sin(2\,\phi_{31})\,.
\end{align}
Notice that $\varepsilon_1$ and $\mathcal{J}_{CP}$ have opposite signs in this
case, once again in agreement with Table~\ref{table2}.

\begin{figure*}
$$\includegraphics[width=10cm]{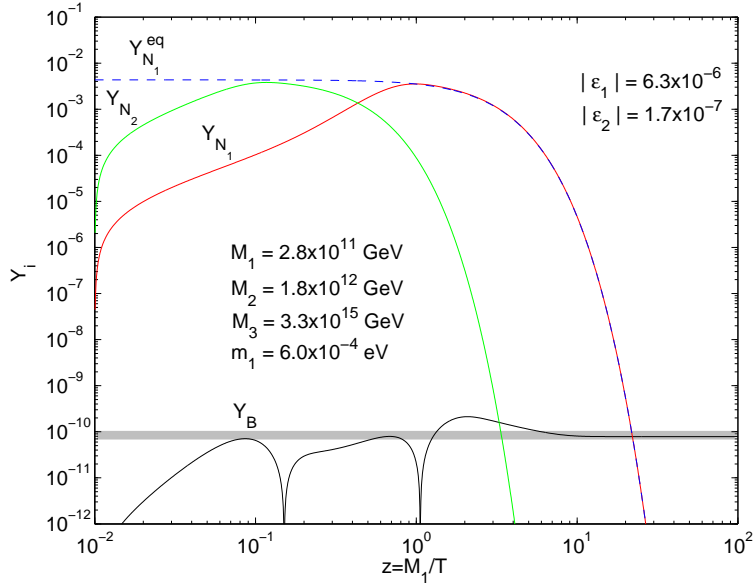}$$
\caption{The same as in Fig.~\ref{fig2} for the example considered in
Eq.~(\ref{YDex2}) with $\phi_{31}=\pi/4$ and $\epsilon=0.4$. The values of the
heavy Majorana neutrino masses $M_i$, the CP asymmetries $\varepsilon_{1,2}$
and the lightest neutrino mass $m_1$ are consistent with the ones presented in
Fig.~\ref{fig3} for this value of $\epsilon$. The predicted value for the final
baryon asymmetry is $Y_B \simeq 8 \times 10^{-11}$. } \label{fig4}
\end{figure*}

In Fig.~\ref{fig3} we present the same numerical analysis as in
Fig.~\ref{fig1}, but for the case where $Y_\nu$ and $D_R$ are defined through
Eqs.~(\ref{YDex2}). We find good agreement between our approximate analytical
results and the numerical ones. The integration of the Boltzmann equations is
plotted in Fig.~\ref{fig4}. The value for the final baryon asymmetry is $Y_B
\simeq 8 \times 10^{-11}$.
\subsection{Two-fold quasi-degeneracy $\bm{(M_1 \simeq M_2 \ll M_3)}$}
\label{degeneracy}%

As an example, let us consider the texture of type I given in
Table~\ref{table1} and assume that $M_1 \simeq M_2 \ll M_3$. The Hermitian
matrix $H_\triangle =Y^\dag_\triangle Y_{\triangle}^{}$ is given by
\begin{align}\label{HDex3}
H_\triangle=\left(\begin{array}{ccc} y_{11}^2+y_{21}^2
&y_{21}\,y_{22}\,e^{-i\phi_{21}}
&\quad 0 \\
y_{21}\,y_{22}\,e^{i\phi_{21}}   & y_{22}^2+y_{32}^2
&y_{32}\,y_{33}\,e^{-i\phi_{32}} \\
0 &y_{32}\,y_{33}\,e^{i\phi_{32}} &y_{33}^2
\end{array}\right)\,.
\end{align}

Taking into account the requirement of Eq.~(\ref{dNlim}) we can get the range
of validity of the parameter $\delta_N$:
\begin{align}
\label{dNex3} \delta_N \gg \frac{1}{8 \pi} {\rm max}
\left\{y_{11}^2+y_{21}^2\;,\; y_{22}^2 +y_{32}^2\right\}\,.
\end{align}
Using Eqs.~(\ref{e12d2}), the CP asymmetries are given in this case by
\begin{align}
\label{esex3} \varepsilon_1 \simeq
\frac{y_{21}^2\,y_{22}^2\,\sin(2\,\phi_{21})}{16\pi\,
\delta_N\,(\,y_{11}^2+y_{21}^2\, )}\;\;,\;\; \varepsilon_2 &\simeq
\frac{1}{16\,\pi}\left[\frac{y_{21}^2\,y_{22}^2\,\sin(2\,\phi_{21})}{
\delta_N\,(\,y_{22}^2+y_{32}^2\,
)}+\frac{3\,y_{32}^2\,y_{33}^2}{y_{22}^2+y_{32}^2}\,\frac{M_2}{M_3}\sin(2\,
\phi_{32})\right]\,.
\end{align}

Let us now consider the following simple realization of $Y_\nu$ and $D_R$:
\begin{align}\label{YDex3}
Y_\nu =\frac{\Lambda_D}{v}\left(\begin{array}{ccc}
 \epsilon^{10}                    &\quad 0               &\quad 0 \\
 \epsilon^{10}\,e^{i\,\phi_{21}}   &\quad \epsilon^9      &\quad 0 \\
 0                             &\quad \epsilon^9      &\quad 1
\end{array}\right)\;\;,\;\;D_R=\Lambda_R\,{\rm
diag}\,(\epsilon^{20},(1+{\delta_N})\,\epsilon^{20},1)\,,
\end{align}
with $\delta_N \ll 1$. From Eq.~(\ref{dNex3}) we get in this case for
$\epsilon=0.3$:
\begin{align}
\label{dNrng} \delta_N \gg
\frac{\Lambda_D^2\,\epsilon^{18}}{4\,\pi\,v^2} \simeq 10^{-11}\,.
\end{align}
The effective neutrino mass matrix is now given by
\begin{align}
\label{mnex3} \mathcal{M}_\nu=\frac{\Lambda_D^2}{\epsilon^2\,(1+\delta_N)
\Lambda_R}\left(\begin{array}{ccc} \epsilon^2\,(1+\delta_N) & \quad
\epsilon^2\,(1+\delta_N)\,e^{i\,\phi_{21}}
&\quad0 \\
\epsilon^2\,(1+\delta_N)\,e^{i\,\phi_{21}}   & \quad 1+
\epsilon^2\,(1+\delta_N)\,e^{2i\,\phi_{21}}   &\quad 1 \\
 0                                             & \quad 1
 &\quad 1+\epsilon^2\,(1+\delta_N)
\end{array}\right)\,.
\end{align}
\begin{figure*}
$$\includegraphics[width=10cm]{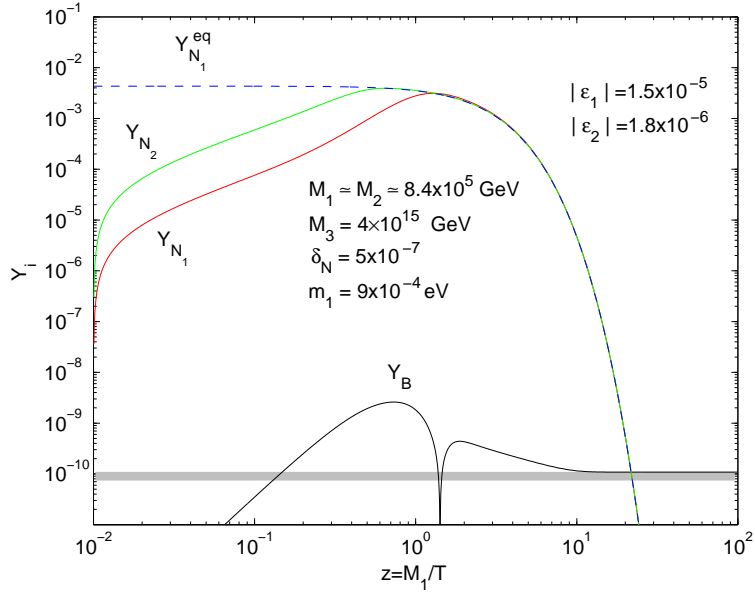}$$
\caption{The same as Figs.~\ref{fig2} and \ref{fig4} for the quasi-degenerate
case $M_1 \simeq M_2$ considered in Eq.~(\ref{YDex3}) with $\phi_{21}=\pi/4$,
$\epsilon=0.3$ and $\delta_N=5 \times 10^{-7}$. The predicted value of the
baryon asymmetry is $Y_B \simeq 10^{-10}$.} \label{fig5}
\end{figure*}
In the limit $\delta_N \rightarrow 0$, Eq.~(\ref{mnex1}) is recovered. Since
$\delta_N \ll 1$ the results for the neutrino masses and mixing parameters are
practically the same as the ones given in Eqs.~(\ref{msex1})-(\ref{angl}). The
same is expected for the estimate of the CP invariant $\mathcal{J}_{CP}$
defined in Eq.~(\ref{esex3}). The main differences reside obviously in the
heavy Majorana neutrino mass spectrum and on the values of the CP asymmetries,
since now the relation $M_1 \ll M_2$ is no longer valid. From Eq.~(\ref{e12d2})
we obtain
\begin{align}
\label{esdeg} \varepsilon_1 &\simeq
\frac{\epsilon^{18}\,\Lambda_D^2\,\sin(2\,\phi_{21})}{32\,\pi\,\delta_N\,v^2}
\simeq \frac{1.3\times 10^{-12}}{\delta_N}\, \sin(2\,\phi_{21})
\,,\nl
\medskip \varepsilon_2 &\simeq\frac{\epsilon^{20}\, \Lambda_D^2\,\sin(2\,
\phi_{21})}{32\,\pi\,\delta_N\,v^2} \simeq \frac{1.1\times
10^{-13}}{\delta_N}\,\sin(2\,\phi_{21})\,.
\end{align}
The results of the numerical integration of the Boltzmann equations for this
case are presented in Fig.~\ref{fig5}. A realistic value for $Y_B$ is also
generated in this case.

Before we end this section, it is worthwhile to comment on the possible effects
of quantum corrections to the effective neutrino mass matrix $\mathcal{M}_\nu$.
This discussion turns out to be relevant since in the examples considered above
we have taken the effective neutrino mass matrix at the heavy neutrino
decoupling scale. Although a detailed treatment would require a renormalization
group analysis for the effective neutrino mass operator, one can employ the
simple analytical treatment considered by many authors in the literature (see
for example Refs.~\cite{GonzalezFelipe:2001kr,Casas:1999tp}). Following this,
we recall that the effective neutrino mass matrix at the electroweak scale
$m_Z$ can be related to the one at the heavy neutrino decoupling scale
$\Lambda_R$ as
\begin{align}
\label{rge1} \mathcal{M}_\nu(m_Z) \propto {\rm
diag}(1,1,1+\epsilon_\tau)\,\mathcal{M}_\nu(\Lambda_R)\,{\rm
diag}(1,1,1+\epsilon_\tau)\,,
\end{align}
where, neglecting the running of the charged lepton Yukawa couplings,
\begin{align}
\label{rge2} \epsilon_\tau \simeq
\frac{3\,y_\tau^2}{32\pi^2}\ln\!\left(\frac{\Lambda_R}{m_Z}\right)\,.
\end{align}
Taking $y_\tau=m_\tau/v$ and $\Lambda_R \lesssim 10^{16}$ GeV, we obtain
$\epsilon_\tau \lesssim 10^{-5}$. Considering $\mathcal{M}_\nu(\Lambda_R)$ as
given in Eq.~(\ref{mnex1}) we would get from Eq.~(\ref{rge1}):
\begin{align}
\label{rge3} &\frac{\dmtwon}
 {\dmthtw}\simeq
 \frac{(1-\epsilon_\tau)\,\epsilon^4}{\sqrt{2}}\quad ,\quad \tan^2\theta_{12}
 \simeq 1-\frac{\epsilon^2}{2\sqrt{2}}-\sqrt{2}\,\epsilon_\tau
 \,,\nonumber \\
&\tan^2\theta_{23}\simeq 1-2\,\epsilon_\tau\quad,\quad |U_{e3}|
\simeq\frac{\epsilon^2}{2\,\sqrt{2}}\,\left(1-\frac{3}{2}
 \,\epsilon_\tau\right).
\end{align}
Since $\epsilon_\tau \ll 1$, the results given by Eqs.~(\ref{rsol}) and
(\ref{angl}) are not significantly altered. The same conclusions are drawn for
the example considered in Eq.~(\ref{mnex2}). Thus, we conclude that the effects
of quantum corrections due to the renormalization of the neutrino mass operator
can be safely neglected in our case.

\section{Concluding remarks} \label{conclusion}

We have analyzed, in the context of the minimal seesaw mechanism, the link
between leptogenesis and CP violation at low energies. In particular, it was
shown that, in order to present a thorough discussion on this question, it is
convenient to work in the WB where both the charged lepton and right-handed
Majorana neutrino mass matrices are diagonal and real, and to write, without
loss of generality, the Dirac neutrino mass matrix as the product of a unitary
matrix and a lower triangular matrix. From the analysis of the phases that
contribute to leptogenesis and low-energy CP violation, we have identified a
necessary condition which is required in order to establish a link between
these two phenomena. We have studied a class of models which satisfy the above
necessary condition in the simplest way, namely those where the Dirac neutrino
mass matrix is of the triangular form. By choosing this structure the number of
physical parameter in the theory is reduced then enhancing its predictability.
In this case there are only three CP-violating phases which contribute both to
leptogenesis and CP violation at low energies. We have then studied the minimal
scenarios where a correct value of the baryon-to-entropy ratio can be
generated, while accounting for all the low-energy neutrino data in the context
of the LMA solution. Moreover, the examples considered in Section~\ref{example}
predict the existence of low-energy CP-violating effects within the range of
sensitivity of the future long-baseline neutrino oscillation experiments. In
fact, it is a remarkable feature of these scenarios that the solutions viable
for leptogenesis are precisely those which maximize $\mathcal{J}_{CP}$. The
question of relating the observed sign of the baryon asymmetry to the sign of
the leptonic CP violation, measurable at low energies through neutrino
oscillations, was also considered. Namely, we have concluded that, within the
minimal scenarios presented, this relation crucially depends on the heavy
Majorana neutrino mass spectrum. We remark that a full discussion of this
aspect requires the computation of the BAU since, besides the prediction of the
relative sign between the BAU and $\mathcal{J}_{CP}$, the determination of
$Y_B$ is of extreme importance to infer about the viability of a given model.

\begin{acknowledgements}
We are grateful to Sacha Davidson for useful discussions. M.N.R. is grateful to
P. Frampton and T. Yanagida for interesting conversations. This work was
partially supported by {\em Funda\c{c}{\~a}o para a Ci{\^e}ncia e a Tecnologia} (FCT,
Portugal) through the projects CERN/FIS/43793/2001, POCTI/36288/FIS/2000,
CERN/FIS/40139/2000 and CFIF - Plurianual (2/91) as well as by the RTN European
Program HPRN-CT-2000-00148 (``AEFNET"). The work of R.G.F. and F.R.J. was
supported by FCT under the grants SFRH/BPD/1549/2000 and
\mbox{PRAXISXXI/BD/18219/98}, respectively.
\end{acknowledgements}

\appendix*
\section{Boltzmann equations}
\label{App1}

The computation of the cosmological baryon asymmetry $Y_B$ involves the
solution of the full set of Boltzmann equations which governs the time
evolution of the right-handed neutrino number densities $Y_{N_j}$ and the
generated lepton asymmetry $Y_{L}$. These quantities depend not only on the
physics occurring in the thermal bath but also on the universe expansion. In
the SM framework extended with heavy right-handed neutrinos the physical
processes relevant to the generation of the baryon asymmetry are typically the
$N_i$ decays and inverse decays into Higgs bosons and leptons, the
$\Delta\,L=1$ scatterings involving the top quark and $\Delta\,L=2$ scatterings
with virtual $N_i$ and $N_i-N_j$ scatterings \cite{Luty:un,Plumacher:1997ru}.
The initially produced lepton asymmetry $Y_L$ is converted into a net baryon
asymmetry $Y_B$ through the $(B+L)$-violating sphaleron processes. One finds
the relation
\begin{align} \label{YBYL}
Y_B=\xi\,Y_{B-L}=\frac{\xi}{\xi-1}\,Y_L\,.
\end{align}
For the SM with three heavy Majorana neutrinos $\xi \simeq 1/3$ and therefore
$Y_B \simeq Y_{B-L}/3 \simeq -Y_L/2\,$.

The Boltzmann equations for the $N_i$ number densities and the $Y_{B-L}$
asymmetry in terms of the dimensionless parameter $z=M_1/T$, where $T$ is the
temperature, can be written in the form:
\begin{align}
 \label{boltz} \frac{dY_{N_j}}{dz} & =-\frac{z}{H s(z)}\left[
\left(\gamma_j^{D}+\gamma_j^{Q}\right)\left( \frac{Y_{N_j}}{Y^{\rm
eq}_{N_j}}-1\right)+\sum_{i=1}^{3}\gamma^{NN}_{ij}\left(
\frac{Y_{N_i}Y_{N_j}}{Y^{\rm eq}_{N_i}Y^{\rm eq}_{N_j}}-1\right)
\right]\ , \nl \noalign{\medskip}
 \frac{dY_{B-L}}{dz} & =-\frac{z}{H
 s(z)}\left[\sum_{j=1}^{3}\varepsilon_j\,
\gamma_j^{D}\left( \frac{Y_{N_j}}{Y^{\rm eq}_{N_j}}-1
\right)+\gamma^{W} \frac{Y_{B-L}}{Y^{\rm eq}_\ell} \right]\ ,
\end{align}
where $H$ is the Hubble parameter evaluated at $z=1$ and $s(z)$ is the entropy
density given by
\begin{align}
\label{Hs} H = \sqrt{\frac{4 \pi^3 g_{\ast}}{45}}\,
\frac{M_1^2}{M_P}\quad,\quad s(z) =\frac{2 \pi^2 g_{\ast}}{45}
\frac{M_1^3}{z^3}\ ,
\end{align}
respectively. Here, $g_{\ast}\simeq106.75$ is the effective number of
relativistic degrees of freedom and $M_P \simeq 1.2\times 10^{19}$ GeV is the
Planck mass. The equilibrium number density of a particle $i$ with mass $m_i$
is given by
\begin{align}
\label{Yeq} Y^{\rm eq}_i(z)=\frac{45}{4\pi^4}\frac{g_i}{g_{\ast}}
\left(\frac{m_i}{M_1}\right)^2 z^2
K_2\left(\frac{m_i\,z}{M_1}\right)\,,
\end{align}
where $g_i$ denote the internal degrees of freedom of the corresponding
particle ($g_{N_i}=2$, $g_\ell=4$) and $K_n(x)$ are the modified Bessel
functions. The $\gamma$'s are the reaction densities for the different
processes. For the decays one has
\begin{align}
\label{gmdec} \gamma_j^{D}&=\frac{M_1 M_j^3}{8
\pi^3z}\left(Y_\nu^\dag Y_\nu\right)_{jj}K_1\!\left(\frac{z
M_j}{M_1}\right)\,.\end{align}

The reaction densities for the $\Delta L=1$ processes with the top quark and
for the $\Delta L=2$ scatterings $N_i-N_j$ can be written in the following way:
\begin{align}
\label{gmQNN}
\gamma_j^{Q}&=2\gamma_{t_j}^{(1)}+4\gamma_{t_j}^{(2)}\quad,\quad
\gamma_{ij}^{NN}=\gamma_{N_i N_j}^{(1)}+\gamma_{N_i N_j}^{(2)}\,,
\end{align}
respectively. Finally, $\gamma^W$ accounts for the washout processes:
\begin{align}
\label{gamW}
\gamma^{W}&=\sum_{j=1}^{3}\left(\frac{1}{2}\gamma_j^{D}+\frac{Y_{N_j}}{Y^{\rm
eq}_{N_j}}\gamma_{t_j}^{(1)}+2\gamma_{t_j}^{(2)}\right)+2\,\gamma_N^{(1)}+2\,
\gamma_N^{(2)}\,.
\end{align}
Each of the above reaction densities can be computed through the corresponding
reduced cross section $\hat{\sigma}^{(i)}(x)$:
\begin{align}
\label{gdef} \gamma^{(i)}(z)=\frac{M_1^4}{64\,\pi^4 z}
\!\!\!\!\operatornamewithlimits{\int}_{(m_a^2+m_b^2)/M_1^2}^{\;\;\;\infty}
\!\!\!\!\hat{\sigma}^{(i)}(x)\,\sqrt{x}\,K_1(z\,\sqrt{x})\,dx\,,
\end{align}
where $x=s/M_1^2\ $, with $s$ being the center-of-mass energy squared and
$m_{a,b}$ are the masses of the initial particle states. All the relevant
reduced cross sections are summarized below. For a more detailed presentation
the reader is addressed to Ref.~\cite{Plumacher:1997ru}.

We write the reduced cross sections as functions of the parameter $x$, the
Hermitian matrix $H_\nu=Y_\nu^\dag Y_\nu$ and the quantities $a_j$ and $c_j$
defined as
\begin{align}
\label{Aaj} a_j=\left(\frac{M_j}{M_1}\right)^2\quad,\quad c_j=
\left(\frac{\Gamma_j}{M_1}\right)^2 =
\frac{a_j\,(H_\nu)_{jj}^2}{64\, \pi^2}\,,
\end{align}
where $\Gamma_j$ is the decay rate defined in Eq.~(\ref{crit1}).

For the $\Delta L=1$ processes involving interactions with quarks we have
\begin{align}
\label{Atj1} \hat{\sigma}_{t_j}^{(1)}(x)&\equiv
\hat{\sigma}(N_j+\ell\leftrightarrow
q+\bar{u})=3\,\alpha_u\,(H_\nu)_{jj}\left(\frac{x-a_j}{x}\right)^2\,,\\
\noalign{\medskip} \hat{\sigma}_{t_j}^{(2)}(x)& \equiv
\hat{\sigma}(N_j+u\leftrightarrow
\bar{\ell}+q)=\hat{\sigma}(N_j+\bar{q}
\leftrightarrow \bar{\ell}+\bar{u}) \nonumber \\
&=3\,\alpha_u\,(H_\nu)_{jj}\,\frac{x-a_j}{x}\,\left[\,1+\frac{a_h-a_j}
{x-a_j+a_h}+\frac{a_j-2
a_h}{x-a_j}\ln\left(\frac {x-a_j+a_h}{a_h}\right) \right]\,,
\end{align}
where $a_h$ (introduced to regularize the infrared divergencies) and
$\alpha_u$ are given by
\begin{align}
\label{Aahau} a_h=\left(\frac{\mu}{M_1}\right)^2\,,\quad
\alpha_u=\frac{{\rm Tr}(Y_u^\dag Y_u^{})}{4\pi}\simeq \frac{m_t^2}{4\pi
v^2}\,,
\end{align}
respectively. The mass parameter $\mu$ is chosen to be $\mu=800$~GeV as in
Refs.~\cite{Luty:un,Plumacher:1997ru}.

The reduced cross sections relative to the $N_i-N_j$ scatterings are:
\begin{align}
\label{ANiNj} \hat{\sigma}_{N_i N_j}^{(1)}&\equiv
\hat{\sigma}(N_i+N_j \leftrightarrow \ell+\bar{\ell})
\nl&=\frac{1}{2\pi}\left\{\,(H_\nu)_{ii}(H_\nu)_{jj}\left[\,\frac{\sqrt{\lambda_{ij}}}{x}
+\frac{a_i+a_j}{2x}\,L_{ij}\,\right]-{\rm
Re}\left[(H_\nu)_{ji}^2\right]
\frac{\sqrt{a_i a_j}}{x-a_i-a_j}\,L_{ij}\right\} \,,\\
\hat{\sigma}_{N_i N_j}^{(2)}&\equiv \hat{\sigma}(N_i+N_j
\leftrightarrow \phi+\phi^\dagger)\nl
&=\frac{1}{2\pi}\left\{\left|\,(H_\nu)_{ji}\right|^{\,2}\left[\,
\frac{L_{ij}}{2}-\frac{\sqrt{\lambda_{ij}}}{x} \,\right]-{\rm
Re}\left[(H_\nu)_{ji}^2\right] \frac{\sqrt{a_i
a_j}\,(a_i+a_j)}{x\,(x-a_i-a_j)}\,L_{ij}\right\}\,,
\end{align}
where the functions $L_{ij}$ and $\lambda_{ij}$ read as:
\begin{align}
\label{Lij} L_{ij}=2
\ln\!\left(\frac{|x+\sqrt{\lambda_{ij}}-a_i-a_j|}{2\,\sqrt{a_i
a_j}}\right)\;\;,\;\;\lambda_{ij}=x^2-2x\,(a_i+a_j)+(a_i-a_j)^2\,.
\end{align}
Since the heavy Majorana neutrinos are unstable particles, one has to replace
the usual fermion propagators by the off-shell propagators $D_j(x)$ in the
computation of the amplitudes for the right-handed neutrino mediated processes:
\begin{align}
\label{AD}
\frac{1}{D_j}\equiv\frac{1}{D_j(x)}=\frac{x-a_j}{(x-a_j)^2+a_jc_j}\,.
\end{align}
For the $\Delta L= 2$ processes the reduced cross sections read then
\begin{align}
\label{AN1N2} \hat{\sigma}_N^{(1)}&\equiv \hat{\sigma}(\ell+\phi
\leftrightarrow
\bar{\ell}+\phi^\dagger)=\sum_{j=1}^{3}\,\left(H_\nu\right)_{jj}^2
\mathcal{A}_{jj}^{(1)}+ \sum_{\overset{n,j=1}{j<n}}^{3}{\rm
Re}\left[\left(H_\nu\right)_{nj}^2\right]\mathcal{B}_{nj}^{(1)}\,,
\nl \hat{\sigma}_N^{(2)}& \equiv \hat{\sigma}(\ell+\ell
\leftrightarrow
\phi^\dagger+\phi^\dagger)=\sum_{j=1}^{3}\,\left(H_\nu\right)_{jj}^2
\mathcal{A}_{jj}^{(2)}+ \sum_{\overset{n,j=1}{j<n}}^{3}{\rm
Re}\left[\left(H_\nu\right)_{nj}^2\right]\mathcal{B}_{nj}^{(2)}\,,
\end{align}
where
\begin{align}
\label{ACA}
\mathcal{A}_{jj}^{(1)}&=\frac{1}{2\pi}\left[1+\frac{a_j}{D_j}+\frac{a_j
\,x}{2\,D_j^2}-\frac{a_j}{x}\left(1+\frac{x+a_j}{D_j}\right)
\ln\left(\frac{x+a_j}{a_j}\right)\right]\,,\nl\noalign{\medskip}
\mathcal{B}_{nj}^{(1)}&=\frac{\sqrt{a_n
a_j}}{2\pi}\left[\frac{1}{D_j}+\frac{1}{D_n}+\frac{x}{D_jD_n}+\left(1+\frac{a_j}
{x}\right)\left(\frac{2}{a_n-a_j}-\frac{1}{D_n}\right)
\ln\left(\frac{x+a_j}{a_j}\right)\right.\nl &
\left.+\left(1+\frac{a_n}{x}\right)\left(
\frac{2}{a_j-a_n}-\frac{1}{D_j}\right)
\ln\left(\frac{x+a_n}{a_n}\right)\right]\,,\nl\noalign{\medskip}
\mathcal{A}_{jj}^{(2)}&=\frac{1}{2\pi}\left[\frac{x}{x+a_j}+\frac{a_j}{x+2a_j}
\ln\left(\frac{x+a_j}{a_j}\right)\right]\,,\nl\noalign{\medskip}
\mathcal{B}_{nj}^{(2)}&=\frac{\sqrt{a_n
a_j}}{2\pi}\left\{\frac{1}{x+a_n+a_j}\,\ln\left[\frac{(x+a_j)(x+a_n)}{a_j\,
a_n}\right] + \frac{2}{a_n-a_j}\,
\ln\left[\frac{a_n\,(x+a_j)}{a_j\,(x+a_n)}\right]\right\}\,.
\end{align}
In our analysis we have computed numerically the reaction densities through
Eq.~(\ref{gdef}) and the above definitions of the reduced cross sections.
Nevertheless, useful analytical approximations can be obtained for specific
temperature regimes \cite{Plumacher:1997ru}.


\begin{thebibliography}{99}
\bibitem{Fukuda:2001nj}
S.~Fukuda {\it et al.}  [Super-Kamiokande Collaboration],
%``Solar B-8 and he p neutrino measurements from 1258 days of
%Super-Kamiokande data,''
Phys.\ Rev.\ Lett.\  {\bf 86}, 5651 (2001) [arXiv:hep-ex/0103032];
%%CITATION = HEP-EX 0103032;%%
S.~Fukuda {\it et al.}  [Super-Kamiokande Collaboration],
%``Constraints on neutrino oscillations using 1258 days of
%Super-Kamiokande solar neutrino data,''
Phys.\ Rev.\ Lett.\  {\bf 86}, 5656 (2001) [arXiv:hep-ex/0103033].
%%CITATION = HEP-EX 0103033;%%

\bibitem{Ahmad:2001an}
Q.~R.~Ahmad {\it et al.}  [SNO Collaboration],
%``Measurement of the charged current interactions produced by B-8  solar
%neutrinos at the Sudbury Neutrino Observatory,''
Phys.\ Rev.\ Lett.\  {\bf 87}, 071301 (2001)
[arXiv:nucl-ex/0106015].
%%CITATION = NUCL-EX 0106015;%%

\bibitem{Masina:2001pp}
For recent reviews and a set of references on neutrino mass models
see, \emph{e.g.}, I.~Masina,
%``The problem of neutrino masses in extensions of the standard model,''
Int.\ J.\ Mod.\ Phys.\ A {\bf 16}, 5101 (2001)
[arXiv:hep-ph/0107220];  G.~Altarelli and F.~Feruglio,
%``Theoretical models of neutrino masses and mixings,''
arXiv:hep-ph/0206077; S.~F.~King,
%``Neutrino mass models,''
arXiv:hep-ph/0208266.
%%CITATION = HEP-PH 0107220;%%
%%CITATION = HEP-PH 0206077;%%
%%CITATION = HEP-PH 0208266;%%

\bibitem{Yanagida:1979}
T.~Yanagida, in Proc. of the {\it Workshop on the Unified Theory
and Baryon Number in the Universe}, ed. by O.~Sawada and A.~
Sugamoto (KEK report 79-18, 1979), p.95, Tsukuba, Japan;
M.~Gell-Mann, P.~ Ramond and R.~Slansky, in {\it Supergravity},
ed. by P.~van Nieuwenhuizen and D.~Z.~Freedman (North Holland,
Amsterdam, 1979), p.315;
%\bibitem{Mohapatra:1979ia}
R.~N.~Mohapatra and G.~Senjanovic,
%``Neutrino Mass And Spontaneous Parity Nonconservation,''
Phys.\ Rev.\ Lett.\  {\bf 44}, 912 (1980).
%%CITATION = PRLTA,44,912;%%

\bibitem {Jungman:1995bz} G.~Jungman, M.~Kamionkowski, A.~Kosowsky and
D.~N.~Spergel, Phys.\ Rev.\ D \textbf{54}, 1332 (1996);
M.~Zaldarriaga, D.~N.~Spergel and U.~Seljak, Astrophys.\ J.\
\textbf{488}, 1 (1997);  P.~de Bernardis \textit{et al.},
Astrophys.\ J.\ \textbf{564}, 559 (2002);  C.~Pryke,
N.~W.~Halverson, E.~M.~Leitch, J.~Kovac, J.~E.~Carlstrom,
W.~L.~Holzapfel and M.~Dragovan, Astrophys.\ J.\ \textbf{568}, 46
(2002).
%%CITATION = ASTRO-PH 9512139;%%
%%CITATION = ASTRO-PH 9702157;%%
%%CITATION = ASTRO-PH 0105296;%%
%%CITATION = ASTRO-PH 0104490;%%

\bibitem {MAP}
http://map.gsfc.nasa.gov/m\underline{ }mm/ms\underline{
}status.html

\bibitem {PLANCK}
{http://astro.estec.esa.nl/Planck/}

\bibitem{Fukugita:1986hr}
M.~Fukugita and T.~Yanagida,
%``Baryogenesis Without Grand Unification,''
Phys.\ Lett.\ B {\bf 174}, 45 (1986).
%%CITATION = PHLTA,B174,45;%%

\bibitem{Lindner:2002vt}
For recent discussions on the physics potential of future
long-baseline neutrino oscillation experiments see, \emph{e.g.},
%\bibitem{Burguet-Castell:2002qx}
J.~Burguet-Castell, M.~B.~Gavela, J.~J.~Gomez-Cadenas,
P.~Hernandez and O.~Mena,
%``Superbeams plus neutrino factory: The golden path to leptonic CP
%violation,''
arXiv:hep-ph/0207080;
%%CITATION = HEP-PH 0207080;%%
M.~Lindner,
%``The physics potential of future long baseline neutrino oscillation
%experiments,''
arXiv:hep-ph/0209083, and references therein.
%%CITATION = HEP-PH 0209083;%%


\bibitem{Endoh:2000hc}
T.~Endoh, T.~Morozumi, T.~Onogi and A.~Purwanto,
%``CP violation in seesaw model,''
Phys.\ Rev.\ D {\bf 64},  013006 (2001) [Erratum, ibid.\ D {\bf 64},
059904 (2001)] [arXiv:hep-ph/0012345].
%%CITATION = HEP-PH 0012345;%%

\bibitem{Branco:gr}
G.~C.~Branco, L.~Lavoura and M.~N.~Rebelo,
%``Majorana Neutrinos And CP Violation In The Leptonic Sector,''
Phys.\ Lett.\ B {\bf 180}, 264 (1986).
%%CITATION = PHLTA,B180,264;%%

\bibitem{Branco:2001pq}
G.~C.~Branco, T.~Morozumi, B.~M.~Nobre and M.~N.~Rebelo,
%``A bridge between CP violation at low energies and leptogenesis,''
Nucl.\ Phys.\ B {\bf 617}, 475 (2001) [arXiv:hep-ph/0107164].
%%CITATION = HEP-PH 0107164;%%

\bibitem{Hagiwara:pw}
K.~Hagiwara {\it et al.}  [Particle Data Group Collaboration],
%``Review Of Particle Physics,''
Phys.\ Rev.\ D {\bf 66}, 010001 (2002).
%%CITATION = PHRVA,D66,010001;%%

\bibitem{Branco:1999bw}
G.~C.~Branco, M.~N.~Rebelo and J.~I.~Silva-Marcos,
%``Degenerate and quasi degenerate Majorana neutrinos,''
Phys.\ Rev.\ Lett.\  {\bf 82}, 683 (1999)[hep-ph/9810328].
%%CITATION = HEP-PH 9810328;%%

\bibitem{Covi:1996wh}
L.~Covi, E.~Roulet and F.~Vissani,
%``CP violating decays in leptogenesis scenarios,''
Phys.\ Lett.\ B {\bf 384}, 169 (1996) [arXiv:hep-ph/9605319];
%%CITATION = HEP-PH 9605319;%%
%\bibitem{Buchmuller:1997yu}
W.~Buchm\"uller and M.~Pl\"umacher,
%``CP asymmetry in Majorana neutrino decays,''
Phys.\ Lett.\ B {\bf 431}, 354 (1998) [arXiv:hep-ph/9710460].
%%CITATION = HEP-PH 9710460;%%

\bibitem{Buchmuller:2000nd}
W.~Buchm\"uller and S.~Fredenhagen,
%``Quantum mechanics of baryogenesis,''
Phys.\ Lett.\ B {\bf 483}, 217 (2000) [arXiv:hep-ph/0004145].
%%CITATION = HEP-PH 0004145;%%

\bibitem{Davidson:2002qv}
S.~Davidson and A.~Ibarra,
%``A lower bound on the right-handed neutrino mass from leptogenesis,''
Phys.\ Lett.\ B {\bf 535}, 25 (2002) [arXiv:hep-ph/0202239].
%%CITATION = HEP-PH 0202239;%%

\bibitem{Ellis:2002eh}
J.~R.~Ellis, M.~Raidal and T.~Yanagida,
%``Observable consequences of partially degenerate leptogenesis,''
Phys.\ Lett.\ B {\bf 546}, 228 (2002) [arXiv:hep-ph/0206300].
%%CITATION = HEP-PH 0206300;%%

\bibitem{Ellis:1984eq}
J.~R.~Ellis, J.~E.~Kim and D.~V.~Nanopoulos,
%``Cosmological Gravitino Regeneration And Decay,''
Phys.\ Lett.\ B {\bf 145}, 181 (1984);
%%CITATION = PHLTA,B145,181;%%
%\bibitem{Ellis:1984er}
J.~R.~Ellis, D.~V.~Nanopoulos and S.~Sarkar,
%``The Cosmology Of Decaying Gravitinos,''
Nucl.\ Phys.\ B {\bf 259}, 175 (1985);
%%CITATION = NUPHA,B259,175;%%
%\bibitem{Moroi:1993mb}
T.~Moroi, H.~Murayama and M.~Yamaguchi,
%``Cosmological constraints on the light stable gravitino,''
Phys.\ Lett.\ B {\bf 303}, 289 (1993);
%%CITATION = PHLTA,B303,289;%%
%\bibitem{Kawasaki:1994af}
M.~Kawasaki and T.~Moroi,
%``Gravitino production in the inflationary universe and the effects on big
%bang nucleosynthesis,''
Prog.\ Theor.\ Phys.\  {\bf 93}, 879 (1995)
[arXiv:hep-ph/9403364];
%%CITATION = HEP-PH 9403364;%%
%\bibitem{Bolz:2000fu}
M.~Bolz, A.~Brandenburg and W.~Buchm\"uller,
%``Thermal production of gravitinos,''
Nucl.\ Phys.\ B {\bf 606}, 518 (2001) [arXiv:hep-ph/0012052].
%%CITATION = HEP-PH 0012052;%%

\bibitem{Giudice:1999fb}
G.~F.~Giudice, M.~Peloso, A.~Riotto and I.~Tkachev,
%``Production of massive fermions at preheating and leptogenesis,''
JHEP {\bf 9908}, 014 (1999) [arXiv:hep-ph/9905242];
%%CITATION = HEP-PH 9905242;%%
%\bibitem{Asaka:1999yd}
T.~Asaka, K.~Hamaguchi, M.~Kawasaki and T.~Yanagida,
%``Leptogenesis in inflaton decay,''
Phys.\ Lett.\ B {\bf 464}, 12 (1999) [arXiv:hep-ph/9906366];
%%CITATION = HEP-PH 9906366;%%
%\bibitem{Asaka:1999jb}
T.~Asaka, K.~Hamaguchi, M.~Kawasaki and T.~Yanagida,
%``Leptogenesis in inflationary universe,''
Phys.\ Rev.\ D {\bf 61}, 083512 (2000) [arXiv:hep-ph/9907559];
%%CITATION = HEP-PH 9907559;%%
%\bibitem{Garcia-Bellido:2001cb}
J.~Garcia-Bellido and E.~Ruiz Morales,
%``Particle production from symmetry breaking after inflation,''
Phys.\ Lett.\ B {\bf 536}, 193 (2002) [arXiv:hep-ph/0109230];
%%CITATION = HEP-PH 0109230;%%
%\bibitem{Fujii:2002jw}
M.~Fujii, K.~Hamaguchi and T.~Yanagida,
%``Leptogenesis with almost degenerate Majorana neutrinos,''
Phys.\ Rev.\ D {\bf 65}, 115012 (2002) [arXiv:hep-ph/0202210];
%%CITATION = HEP-PH 0202210;%%
%\bibitem{Allahverdi:2002gz}
R.~Allahverdi and A.~Mazumdar,
%``Non-thermal leptogenesis with almost degenerate superheavy neutrinos,''
Phys.\ Rev.\ D {\bf 67}, 023509 (2003) [arXiv:hep-ph/0208268].
%%CITATION = HEP-PH 0208268;%%
%\bibitem{Boubekeur:2002gv}
L.~Boubekeur, S.~Davidson, M.~Peloso and L.~Sorbo,
%``Leptogenesis and rescattering in supersymmetric models,''
arXiv:hep-ph/0209256.
%%CITATION = HEP-PH 0209256;%%

\bibitem{Flanz:1994yx}
M.~Flanz, E.~A.~Paschos and U.~Sarkar,
%``Baryogenesis from a lepton asymmetric universe,''
Phys.\ Lett.\ B {\bf 345}, 248 (1995) [Erratum-ibid.\ B {\bf 382},
447 (1996)] [arXiv:hep-ph/9411366];
%%CITATION = HEP-PH 9411366;%%
%\bibitem{Pilaftsis:1997jf}
A.~Pilaftsis,
%``CP violation and baryogenesis due to heavy Majorana neutrinos,''
Phys.\ Rev.\ D {\bf 56}, 5431 (1997) [arXiv:hep-ph/9707235];
%%CITATION = HEP-PH 9707235;%%
%\bibitem{Buchmuller:1997yu}
W.~Buchm\"uller and M.~Pl\"umacher,
%``CP asymmetry in Majorana neutrino decays,''
Phys.\ Lett.\ B {\bf 431}, 354 (1998) [arXiv:hep-ph/9710460];
%%CITATION = HEP-PH 9710460;%%
%\bibitem{Pilaftsis:1998pd}
A.~Pilaftsis,
%``Heavy Majorana neutrinos and baryogenesis,''
Int.\ J.\ Mod.\ Phys.\ A {\bf 14}, 1811 (1999)
[arXiv:hep-ph/9812256].
%%CITATION = HEP-PH 9812256;%%

\bibitem{GonzalezFelipe:2001kr}
R.~Gonz\'alez Felipe and F.~R.~Joaquim,
%``Is right-handed neutrino degeneracy compatible with the solar and
%atmospheric neutrino data?,''
JHEP {\bf 0109}, 015 (2001) [arXiv:hep-ph/0106226].
%%CITATION = HEP-PH 0106226;%%

\bibitem{Branco:2002kt}
G.~C.~Branco, R.~Gonz\'alez Felipe, F.~R.~Joaquim and
M.~N.~Rebelo,
%``Leptogenesis, CP violation and neutrino data: What can we learn?,''
Nucl.\ Phys.\ B {\bf 640}, 202 (2002) [arXiv:hep-ph/0202030].
%%CITATION = HEP-PH 0202030;%%

\bibitem{Frampton:2002qc}
P.~H.~Frampton, S.~L.~Glashow and T.~Yanagida,
%``Cosmological sign of neutrino CP violation,''
Phys.\ Lett.\ B {\bf 548}, 119 (2002) [arXiv:hep-ph/0208157].
%%CITATION = HEP-PH 0208157;%%

\bibitem{Casas:2001sr}
J.~A.~Casas and A.~Ibarra,
%``Oscillating neutrinos and mu $\to$ e, gamma,''
Nucl.\ Phys.\ B {\bf 618}, 171 (2001) [arXiv:hep-ph/0103065].
%%CITATION = HEP-PH 0103065;%%

\bibitem{Lavignac:2002gf}
S.~Lavignac, I.~Masina and C.~A.~Savoy,
%``Large solar angle and seesaw mechanism: A bottom-up perspective,''
Nucl.\ Phys.\ B {\bf 633}, 139 (2002) [arXiv:hep-ph/0202086].
%%CITATION = HEP-PH 0202086;%%

\bibitem{Masina:2002qh}
I.~Masina,
%``Lepton flavour violation,''
arXiv:hep-ph/0210125.
%%CITATION = HEP-PH 0210125;%%

\bibitem{Rebelo:2002wj}
M.~N.~Rebelo,
%``Leptogenesis without CP violation at low energies,''
Phys.\ Rev.\ D {\bf 67}, 013008 (2003) [arXiv:hep-ph/0207236].
%%CITATION = HEP-PH 0207236;%%

\bibitem{Altarelli:1999dg}
G.~Altarelli, F.~Feruglio and I.~Masina,
%``Large neutrino mixing from small quark and lepton mixings,''
Phys.\ Lett.\ B {\bf 472}, 382 (2000) [arXiv:hep-ph/9907532].
%%CITATION = HEP-PH 9907532;%%

\bibitem{Smirnov:af}
A.~Y.~Smirnov,
%``Seesaw Enhancement Of Lepton Mixing,''
Phys.\ Rev.\ D {\bf 48}, 3264 (1993) [arXiv:hep-ph/9304205];
%%CITATION = HEP-PH 9304205;%%
%\bibitem{King:1999mb}
%\bibitem{King:1999cm}
S.~F.~King,
%``Atmospheric and solar neutrinos from single right-handed neutrino
%dominance and U(1) family symmetry,''
Nucl.\ Phys.\ B {\bf 562}, 57 (1999) [arXiv:hep-ph/9904210];
%%CITATION = HEP-PH 9904210;%%
S.~F.~King,
%``Large mixing angle MSW and atmospheric neutrinos from single  right-handed
% neutrino dominance and U(1) family symmetry,''
Nucl.\ Phys.\ B {\bf 576}, 85 (2000) [arXiv:hep-ph/9912492];
%%CITATION = HEP-PH 9912492;%%
%\bibitem{Hirsch:2001dg}
M.~Hirsch and S.~F.~King,
%``Leptogenesis with single right-handed neutrino dominance,''
Phys.\ Rev.\ D {\bf 64}, 113005 (2001) [arXiv:hep-ph/0107014];
%%CITATION = HEP-PH 0107014;%%
%\bibitem{King:2002nf}
S.~F.~King,
%``Constructing the large mixing angle MNS matrix in see-saw models with
%right-handed neutrino dominance,''
JHEP {\bf 0209}, 0411 (2002) [arXiv:hep-ph/0204360].
%%CITATION = HEP-PH 0204360;%%

\bibitem{inprep} Lisbon-Saclay collaboration, in preparation.

\bibitem{Ellis:2002xg}
J.~R.~Ellis and M.~Raidal,
%``Leptogenesis and the violation of lepton number and CP at low energies,''
Nucl.\ Phys.\ B {\bf 643}, 229 (2002) [arXiv:hep-ph/0206174];
%%CITATION = HEP-PH 0206174;%%
%\bibitem{Davidson:2002em}
S.~Davidson and A.~Ibarra,
%``Leptogenesis and low-energy phases,''
Nucl.\ Phys.\ B {\bf 648}, 345 (2003) [arXiv:hep-ph/0206304].
%%CITATION = HEP-PH 0206304;%%

\bibitem{Buchmuller:2001dc}
W.~Buchm\"uller and D.~Wyler,
%``CP violation, neutrino mixing and the baryon asymmetry,''
Phys.\ Lett.\ B {\bf 521}, 291 (2001) [arXiv:hep-ph/0108216];
%%CITATION = HEP-PH 0108216;%%
%\bibitem{Buccella:2001tq}
F.~Buccella, D.~Falcone and F.~Tramontano,
%``Baryogenesis via leptogenesis in SO(10) models,''
Phys.\ Lett.\ B {\bf 524}, 241 (2002) [arXiv:hep-ph/0108172];
%%CITATION = HEP-PH 0108172;%%
%\bibitem{Ellis:2001xt}
J.~R.~Ellis, J.~Hisano, S.~Lola and M.~Raidal,
%``CP violation in the minimal supersymmetric seesaw model,''
Nucl.\ Phys.\ B {\bf 621}, 208 (2002) [arXiv:hep-ph/0109125];
%%CITATION = HEP-PH 0109125;%%
%\bibitem{Rodejohann:2002mh}
W.~Rodejohann and K.~R.~Balaji,
%``Leptogenesis and low energy observables in left-right symmetric models,''
Phys.\ Rev.\ D {\bf 65}, 093009 (2002) [arXiv:hep-ph/0201052];
%%CITATION = HEP-PH 0201052;%%
%\bibitem{Nielsen:2002pc}
H.~B.~Nielsen and Y.~Takanishi,
%``Baryogenesis via lepton number violation and family replicated gauge  group,''
Nucl.\ Phys.\ B {\bf 636}, 305 (2002) [arXiv:hep-ph/0204027];
%%CITATION = HEP-PH 0204027;%%
%\bibitem{Buchmuller:2002rq}
W.~Buchm\"uller, P.~Di Bari and M.~Pl\"umacher,
%``Cosmic microwave background, matter-antimatter asymmetry and neutrino
%masses,''
Nucl.\ Phys.\ B {\bf 643}, 367 (2002) [arXiv:hep-ph/0205349];
%%CITATION = HEP-PH 0205349;%%
%\bibitem{Rodejohann:2002hx}
W.~Rodejohann,
%``Leptogenesis, mass hierarchies and low energy parameters,''
Phys.\ Lett.\ B {\bf 542}, 100 (2002) [arXiv:hep-ph/0207053];
%%CITATION = HEP-PH 0207053;%%
%\bibitem{Endoh:2002wm}
T.~Endoh, S.~Kaneko, S.~K.~Kang, T.~Morozumi and M.~Tanimoto,
%``CP violation in neutrino oscillation and leptogenesis,''
Phys.\ Rev.\ Lett.\  {\bf 89}, 231601 (2002) [arXiv:hep-ph/0209020].
%%CITATION = HEP-PH 0209020;%%
%\bibitem{Xing:2002yq}
Z.~Z.~Xing,
%``Leptogenesis and CP violation in neutrino oscillations,''
arXiv:hep-ph/0209066;
%%CITATION = HEP-PH 0209066;%%
%\bibitem{Pati:2002pe}
J.~C.~Pati,
%``Leptogenesis Within A Predictive G(224)/SO(10)-Framework,''
arXiv:hep-ph/0209160;
%%CITATION = HEP-PH 0209160;%%
%\bibitem{Raidal:2002xf}
M.~Raidal and A.~Strumia,
%``Predictions of the most minimal see-saw model,''
Phys.\ Lett.\ B {\bf 553}, 72 (2003) [arXiv:hep-ph/0210021].
%%CITATION = HEP-PH 0210021;%%
%\bibitem{Hambye:2002nw}
T.~Hambye,
%``Tests of Leptogenesis at Low Energy,''
arXiv:hep-ph/0210048;
%%CITATION = HEP-PH 0210048;%%
%\bibitem{Kaneko:2002yp}
S.~Kaneko and M.~Tanimoto,
%``Neutrino mass matrix with two zeros and leptogenesis,''
Phys.\ Lett.\ B {\bf 551}, 127 (2003) [arXiv:hep-ph/0210155].
%%CITATION = HEP-PH 0210155;%%

\bibitem{Plumacher:1996kc}
M.~Pl\"umacher,
%``Baryogenesis and lepton number violation,''
Z.\ Phys.\ C {\bf 74}, 549 (1997) [arXiv:hep-ph/9604229].
%%CITATION = HEP-PH 9604229;%%

\bibitem{Kageyama:2002zw}
A.~Kageyama, S.~Kaneko, N.~Shimoyama and M.~Tanimoto,
%``See-saw realization of the texture zeros in the neutrino mass matrix,''
Phys.\ Lett.\ B {\bf 538}, 96 (2002) [arXiv:hep-ph/0204291].
%%CITATION = HEP-PH 0204291;%%

\bibitem{Frampton:2002yf}
P.~H.~Frampton, S.~L.~Glashow and D.~Marfatia,
%``Zeroes of the neutrino mass matrix,''
Phys.\ Lett.\ B {\bf 536}, 79 (2002) [arXiv:hep-ph/0201008].
%%CITATION = HEP-PH 0201008;%%

\bibitem{Apollonio:1999ae}
M.~Apollonio {\it et al.}  [CHOOZ Collaboration],
%``Limits on neutrino oscillations from the CHOOZ experiment,''
Phys.\ Lett.\ B {\bf 466}, 415 (1999) [arXiv:hep-ex/9907037].
%%CITATION = HEP-EX 9907037;%%

\bibitem{Fusaoka:1998vc}
H.~Fusaoka and Y.~Koide,
%``Updated estimate of running quark masses,''
Phys.\ Rev.\ D {\bf 57}, 3986 (1998) [arXiv:hep-ph/9712201].
%%CITATION = HEP-PH 9712201;%%

\bibitem{Akhmedov:2000yt}
E.~K.~Akhmedov, G.~C.~Branco, F.~R.~Joaquim and
J.~I.~Silva-Marcos,
%``Neutrino masses and mixing with seesaw mechanism and universal breaking
% of extended democracy,''
Phys.\ Lett.\ B {\bf 498}, 237 (2001) [arXiv:hep-ph/0008010].
%%CITATION = HEP-PH 0008010;%%

\bibitem{Casas:1999tp}
J.~A.~Casas, J.~R.~Espinosa, A.~Ibarra and I.~Navarro,
%``Naturalness of nearly degenerate neutrinos,''
Nucl.\ Phys.\ B {\bf 556}, 3 (1999) [arXiv:hep-ph/9904395];
%%CITATION = HEP-PH 9904395;%%
%\bibitem{Barbieri:1999ak}
R.~Barbieri, G.~G.~Ross and A.~Strumia,
%``Vacuum oscillations of quasi degenerate solar neutrinos,''
JHEP {\bf 9910}, 020 (1999) [arXiv:hep-ph/9906470];
%%CITATION = HEP-PH 9906470;%%
%\bibitem{Haba:1999zg}
N.~Haba, Y.~Matsui and N.~Okamura,
%``Analytic solutions to the RG equations of the neutrino physical
%parameters,''
Prog.\ Theor.\ Phys.\  {\bf 103}, 807 (2000)
[arXiv:hep-ph/9911481].
%%CITATION = HEP-PH 9911481;%%

\bibitem{Luty:un}
M.~A.~Luty,
%``Baryogenesis Via Leptogenesis,''
Phys.\ Rev.\ D {\bf 45}, 455 (1992).
%%CITATION = PHRVA,D45,455;%%

\bibitem{Plumacher:1997ru}
M.~Pl\"umacher,
%``Baryon asymmetry, neutrino mixing and supersymmetric SO(10) unification,''
Nucl.\ Phys.\ B {\bf 530}, 207 (1998) [arXiv:hep-ph/9704231].
%%CITATION = HEP-PH 9704231;%%

\end{thebibliography}
\end{document}